\documentclass[a4paper, twoside, reqno, 12pt, dvips]{amsart}
\RequirePackage{fix-cm} 

\usepackage{fixltx2e}     

\usepackage[english]{babel}
\usepackage[latin1]{inputenc}

\usepackage{eucal}
\usepackage{esint}
\usepackage{amsgen}
\usepackage{amsthm}
\usepackage{xspace}
\usepackage{amssymb}
\usepackage{amsmath}
\usepackage{MnSymbol}
\usepackage{amsfonts}
\usepackage{verbatim}
\usepackage{mathrsfs, dsfont}

\usepackage[nice]{nicefrac}

\usepackage{a4wide}

\usepackage{microtype}
\usepackage[bf, up, hang]{caption}

\usepackage{indentfirst}
\usepackage{graphicx}
\usepackage{subfigure}
\usepackage[section]{placeins}
\usepackage{psfrag}

\usepackage[usenames, dvipsnames, pdf]{pstricks}
\usepackage{epsfig}
\usepackage{pst-grad} 
\usepackage{pst-plot} 

\usepackage{color}
\definecolor{oneblue}{rgb}{0.0, 0.0, 0.85}
\definecolor{darkgrey}{rgb}{0.273, 0.281, 0.30}

\usepackage{xcolor}
\definecolor{lightgray}{gray}{0.9}

\usepackage[colorlinks,
            urlcolor=oneblue,
            linkcolor=oneblue,
            citecolor=oneblue,
            bookmarksopen=false,
            pagebackref]{hyperref}

\usepackage[compact]{titlesec}

\titleformat{\section}{\normalfont\Large\bfseries\sffamily\center\color{darkgrey}}{\thesection.}{0.5em}{}{}
\titleformat{\subsection}{\normalfont\large\bfseries\sffamily\color{darkgrey}}{\thesubsection.}{0.4em}{}{}
\titleformat{\subsubsection}{\normalfont\normalsize\bfseries\sffamily\color{darkgrey}}{\thesubsubsection.}{0.3em}{}{}

\titlespacing*{\section}{1.0em}{1.0em}{0.8em}[0em]
\titlespacing*{\subsection}{1.0em}{1.0em}{0.8em}[0em]
\titlespacing*{\subsubsection}{1.0em}{0.7em}{0.6em}[0em]

\usepackage{titletoc}

\contentsmargin{0.0em}
\dottedcontents{section}[2.5em]{\addvspace{0.45em}\bfseries}{1.0em}{0pc}
\dottedcontents{subsection}[4.5em]{}{2.6em}{0pc}
\dottedcontents{subsubsection}[5.5em]{}{3.0em}{0pc}

\usepackage{fancyhdr}
\usepackage{lastpage}

\newcommand*\Title{Numerical simulation of wave impact}
\newcommand*\Authors{A.~Rafiee, D.~Dutykh et al.}

\usepackage{acronym}

\numberwithin{equation}{section}

\newcommand{\up}[1]{$\,^{\mathrm{\small\textsf{#1}}}$} 

\newcommand{\ud}{\mathrm{d}}

\renewcommand{\div}{\grad\scal}

\newcommand{\scal}{\boldsymbol{\cdot}}
\newcommand{\grad}{\boldsymbol{\nabla}}

\newcommand{\half}{{\textstyle{1\over2}}}

\begin{document}

\title[\Title]{Numerical simulation of wave impact on a rigid wall using a two--phase compressible SPH method}

\author[A.~Rafiee]{Ashkan Rafiee$^*$}
\address{University College Dublin, School of Mathematical Sciences, Belfield, Dublin 4, Ireland}
\email{Ashkan.Rafiee@ucd.ie}
\thanks{$^*$ Corresponding author}

\author[D.~Dutykh]{Denys Dutykh}
\address{University College Dublin, School of Mathematical Sciences, Belfield, Dublin 4, Ireland \and LAMA, UMR 5127 CNRS, Universit\'e de Savoie, Campus Scientifique, 73376 Le Bourget-du-Lac Cedex, France}
\email{Denys.Dutykh@ucd.ie}
\urladdr{http://www.denys-dutykh.com/}

\author[F.~Dias]{Fr\'ed\'eric Dias}
\address{CMLA, ENS Cachan, CNRS, 61 Avenue du Pr\'esident Wilson, F-94230 Cachan, France \and University College Dublin, School of Mathematical Sciences, Belfield, Dublin 4, Ireland}
\email{Frederic.Dias@ucd.ie}

\begin{abstract}
In this paper, an SPH method based on the SPH--ALE formulation is used for modelling two-phase flows with large density ratios and realistic sound speeds. The SPH scheme is further improved to circumvent the tensile instability that may occur in the SPH simulations. The two-phase SPH solver is then used to model a benchmark problem of liquid impact on a rigid wall. The results are compared with an incompressible Level Set solver. Furthermore, a wave impact on a rigid wall with a large entrained air pocket is modelled. The SPH simulation is initialised by the output of a fully non-linear potential flow solver. The pressure distribution, velocity field and impact pressure are then analysed.

\bigskip
\noindent \textbf{\keywordsname:} Smoothed Particle Hydrodynamics; Godunov method; Arbitrary Lagrangian--Eulerian formulation; Multi-Phase Flows; Wave Impact

\end{abstract}

\maketitle
\tableofcontents
\thispagestyle{empty}

\section{Introduction}

\begin{figure}
  \centering
  \includegraphics[width=0.85\textwidth]{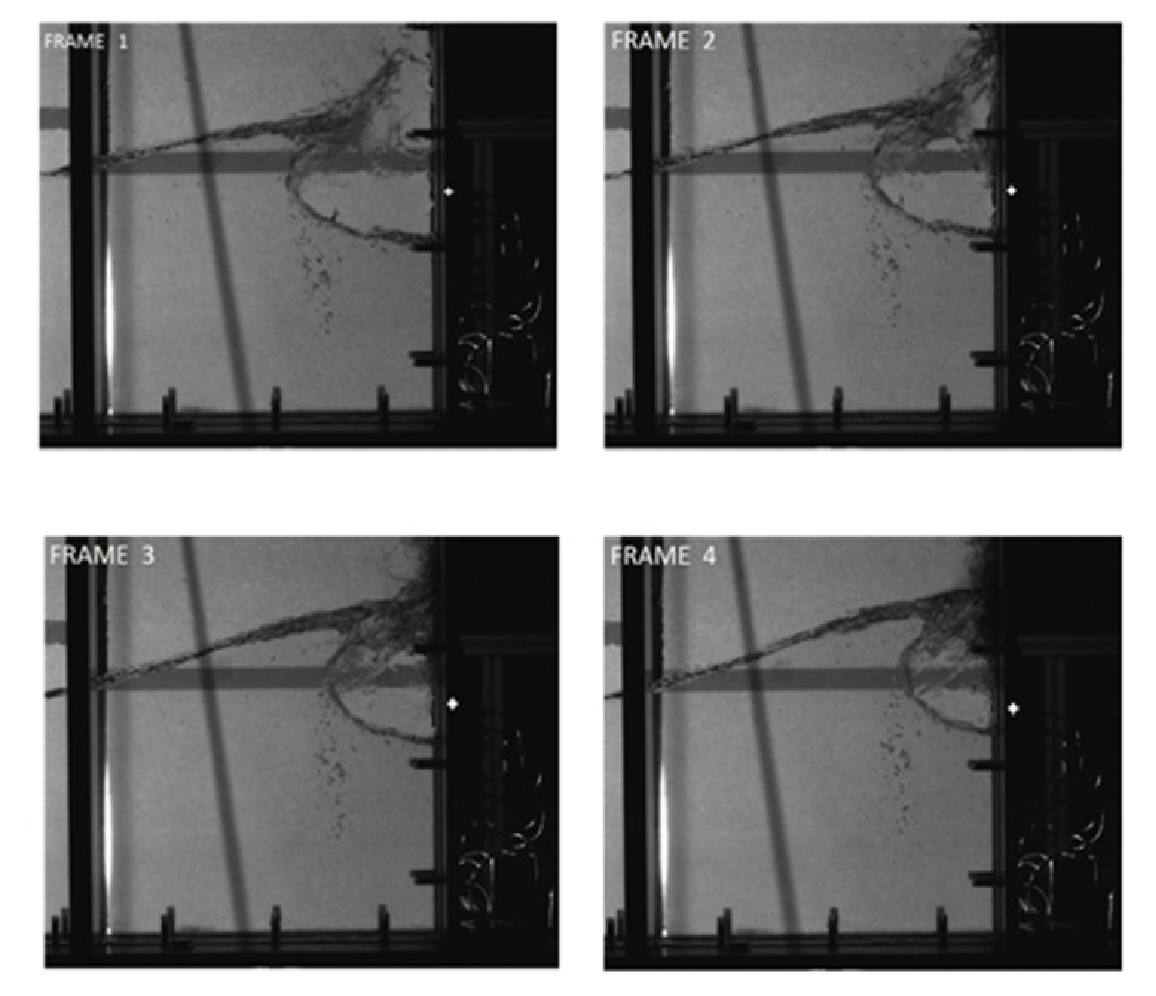}
  \caption{\small\em An air-pocket impact inside a sloshing tank \cite{Repalle2012}.}
  \label{sloshing}
\end{figure}

In many marine engineering applications it is crucial to understand and accurately predict impact forces on the structures. Examples are liquid sloshing inside LNG carriers, wave impact on an offshore platform, wave interaction with a wave energy converter, etc. In the wave--structure interaction context, it is understood that the shape of the impacting wave, the location of the structure relative to the wave's breaking point and the size of the entrapped air pocket have a significant effect on the impact pressure exerted on the structure. An example of wave impact during which the entrapped gaseous phase plays a significant role on the impact pressure is when a large gas pocket is entrapped during the wave impact on a fixed or oscillating structure. Figure~\ref{sloshing} shows an example of air pocket impact in a sloshing tank.

The influence of air during wave impact on structures is a difficult topic and has attracted many researchers. Although it is commonly believed that the presence of air pockets during the impact has a cushioning effect, this may not be the dominant phenomenon \cite{Bredmose1}. It is understood that the shape and size of the entrained air (single pocket or cloud of small bubbles) influences the impact phenomenology \cite{Dias2008}. On the other hand, although the pressure peak may become smaller, the entrapment of air bubbles prolongs the impact duration and also results in pressure oscillations on the wall due to compressions and expansions of the air bubbles. This will consequently increase the pressure impulse on the wall. \textsc{Bagnold} (1939) \cite{Bagnold1939} was the first to realise that the time histories of maximum pressure and impact duration are stochastic and differ from one identical wave impact to another, even in carefully controlled laboratory experiments, while the pressure impulse appears to be more repeatable. \textsc{Wood} \emph{et al.} (2000) \cite{Wood2000} modelled wave impact on a wall with entrapped air using the pressure impulse theory. \textsc{Peregrine} and \textsc{Thais} (1996) \cite{Peregrine1996} studied the influence of entrapped air on violent water wave impacts. \textsc{Bullock} (2001) \cite{Bullock2001} noted that the difference in the properties of air bubbles in salt water compared to the fresh water and observed smaller impact pressures with slat water than with fresh water. \textsc{Bredmose} \emph{et al.} (2009) \cite{Bredmose2009} and \textsc{Dias} \emph{et al.} (2010) \cite{Dias2008} have recently performed numerical studies on violent impact pressure in aerated flows. \textsc{Lugni} \emph{et al.} (2006) \cite{Lugni2006} performed a series of experiments on the influence of gaseous phase on the impact pressure for a flip--through sloshing impact. They studied three different flip--through impacts depending on the amount and size of the entrapped bubbles. They observed that for events when no air was entrapped the after impact pressure decayed almost monotonically, while for events with a single well formed bubble entrainment a distinct oscillation pattern was observed. This was then associated to the rebounding action of the single, well formed entrapped air bubble. Smaller and higher frequency oscillations were also observed during the decay phase of events in which small air bubbles were engulfed and the flow was highly 3D.

A real life liquid impact involves various physical parameters such as the compressibility of the gaseous phase, density ratio between liquid and gaseous phases, speeds of sound of fluids, elasticity of the wall, etc. Therefore, a complete study of liquid impacts is far beyond the capabilities of the current numerical simulations and it can be only studied through delicate model/full scale experiments. Here, we model an air pocket impact using a two-phase compressible SPH code initialised by the output of a fully non-linear potential flow solver. The potential flow solver used is the FSID code developed by \textsc{Y. M. Scolan}. The FSID code is based on the succession of transformal mappings and a desingularised technique \cite{Scolan2010}.

The simulations are carried out using a multi-phase compressible SPH code and an incompressible Level Set code. The Level Set code is based on Finite Differences while the SPH code is based on the SPH--ALE (Arbitrary Lagrangian Eulerian) formulation of the SPH method \cite{Vila1999}. The details of each scheme are presented and results are compared to understand the effect of compressibility/incompressibility on the impact pressure.

\section{Compressible Two-Fluid SPH Model}\label{sec:sph}

The derivation of a multi-fluid SPH scheme  for large density ratios is not trivial and requires a careful calculation of the pressure gradient in the momentum equation. Recently several SPH schemes have been proposed for modelling multi-fluid flows with large density ratios \cite{Hu2006, Grenier2009, Leduc2010, Monaghan2013, Khayyer2012}. All of these multi-fluid SPH schemes have advantages and disadvantages. However, none of these methods is suitable for modelling compressible inviscid two-fluid systems using realistic (physical) speeds of sound of the fluids except the approach of \textsc{Leduc} \emph{et al.} (2010) \cite{Leduc2010} which is based on the SPH--ALE formulation of \cite{Vila1999}. Here, a variation of this SPH scheme will be introduced and used for simulating a compressible inviscid two-fluid system.

Let us consider the system of conservation laws for the Euler equations for each phase \cite{Leduc2010}
\begin{equation}\label{euler_cons}
  L_{v_0} \bigl(\Phi\bigr) + \div\Bigl(F_E\bigl(\Phi \bigr) - v_0\Phi\Bigr) = Q_v
\end{equation}
where, $\Phi = \bigl(\rho, \rho v^{(1)}, \rho v^{(2)} \bigr)^t$ is the vector of conservative variables, $L_v = \partial_t + \sum_{l=1}^{d}v^l\partial_{x^l}$ is the transport operator associated to $v$, $F_E$ is the Eulerian flux matrix and $Q_v$ is the volume source term. \textsc{Vila} \cite{Vila1999} showed that the SPH discretization of Eq.~(\ref{euler_cons}) leads to a one dimensional Riemann problem between each pair of interacting particles. \textsc{Leduc} \emph{et al.} (2010) \cite{Leduc2010} showed that by considering two control volumes on particles $i$ and $j$, Eq.~(\ref{euler_cons}) takes the form
\begin{equation}\label{euler_cons_sph}
\left\{
\begin{split}
&\frac{\partial}{\partial t} \left( \Phi \right) + \frac{\partial}{\partial x_{n_{ij}}} \left( F_E \left( \Phi \right) \cdot n_{ij} - v_0 \left( x_{ij}, t \right) \cdot n_{ij} \Phi  \right) = Q_i \\ \\
&\Phi \left( x_{n_{ij}}, t=0 \right) = \left\{
\begin{split}
\Phi_i,  \quad  \mbox{if}  \quad x_{n_{ij}} < 0 \\
\Phi_j,  \quad  \mbox{if}  \quad x_{n_{ij}} > 0.
\end{split} \right.
\end{split}
\right.
\end{equation}
Here, $n_{ij}$ is the unit vector from $i$ to $j$, $x_{ij}$ is the mid-point between $i$ and $j$ and $x_{n_{ij}}$ is the curvilinear abscissa along the straight line between $i$ and $j$. The final forms of the SPH--ALE equations are then
\begin{equation}
\label{sph_ale}
\left\{
\begin{split}
&\frac{\ud \left( {\bf x}_i \right) }{\ud t} = {\bf v}_0 \left( {\bf x}_i\right) \\
&\frac{\ud \left( \omega _i \right)}{\ud t} = \omega _i \sum _j \omega _j \left( {\bf v}_0 \left( {\bf x}_j \right) -  {\bf v}_0 \left( {\bf x}_i \right) \right) B_{ij} \cdot \nabla _i W_{ij} \\
&\frac{\ud \left( \omega _i \rho _i \right)}{\ud t} + \omega _i \sum _j \omega _j 2 \rho _{E,ij} \left( {\bf v}_{E,ij} - {\bf v}_0 \left( {\bf x}_{ij},t \right) \right) B_{ij} \cdot  \nabla _i W_{ij} = 0\\ 
&\frac{\ud \left( \omega _i \rho _i {\bf v}_i\right) }{\ud t} + \omega _i \sum _j \omega _j 2 [  \rho _{E,ij} {\bf v}_{E,ij} \otimes \left( {\bf v}_{E,ij} - {\bf v}_0 \left( {\bf x}_{ij},t \right) \right) + p_{E,ij} ] B_{ij} \cdot \nabla _i W_{ij} = \omega _i \rho_i {\bf g}
\end{split}
\right.
\end{equation}
where $\bigl(\rho_{E,ij}, v_{E,ij} \bigr)^t = \Phi _{ij}\bigl(\lambda_0^{ij}\bigr)$ is the upwind solution of the moving Riemann problem \cite{Leduc2010} and $B_{ij} = \half\bigl(B_i + B_j\bigr)$ stands for the symetrized renormalization matrix \cite{Vila2005} and takes the form
\begin{equation}\label{sym_ren_vila}
B_i = \left[ \sum _j \omega _j \left( x_j - x_i \right) \otimes \nabla_i W_{ij} \right]^{-1}.
\end{equation}
Here, $\omega_i$ is the volume of particle $i$, $\rho$ is density, $p$ is the pressure, and $W$ and ${\bf g} = 9.81$ $\mathsf{m}\,\mathsf{s}^{-2}$ are the SPH kernel and gravitational acceleration, respectively. Following \cite{Leduc2010}, since the pressure is continuous across the interface of two-fluids, the Riemann problem is solved for the variables $\bigl(p, v^{(1)}, v^{(2)} \bigr)^t$ with the Tait equation of state. Therefore, the vector $\bigl(\rho_{E,ij}, v_{E,ij} \bigr)^t$ in Eq.~(\ref{sph_ale}) is equal to $\bigl(\rho^\ast, v^{(1)\ast}, v^{(2)\ast} \bigr)^t$, where the superscripts $\ast$ denote the solution of the Riemann problem in the star region. The approximated linearized solution of the Riemann problem is given by
\begin{equation}\label{riemann_star}
\left\{
\begin{split}
&{\bf v}^{(1)\ast} = \frac{\rho_l c_l {\bf v}_l ^{(1)} + \rho_r c_r {\bf v}_r ^{(1)}}{\rho_l c_l + \rho_r c_r} - \frac{p_r - p_l}{\rho_l c_l + \rho_r c_r} \\
&p^\ast = \frac{\rho_l c_l p_r + \rho_r c_r p_l}{\rho_l c_l + \rho_r c_r} - \frac{\rho_l c_l \rho_r c_r ( {\bf v}_r ^{(1)} - {\bf v}_l ^{(1)})}{\rho_l c_l + \rho_r c_r}
\end{split}
\right.
\end{equation}
where subscripts $r$ and $l$ denote the right and left states of the Riemann problem and $c$ is the speed of sound. Once $p^\ast$ and ${\bf v}^{(1)\ast}$ are known, one can calculate $\rho^\ast$ with the Tait equation of state from $p^\ast$ and ${\bf v}^{(2)\ast}$ from
\begin{equation}\label{riemann_v2}
\left\{
\begin{split}
&{\bf v}^{(2)\ast} = {\bf v} _l ^{(2)}, \quad \mbox{if} \quad \frac{{\bf x}}{t} < {\bf v}^{(1)\ast} \\
&{\bf v}^{(2)\ast} = {\bf v} _r ^{(2)}, \quad \mbox{otherwise.}
\end{split}
\right.
\end{equation}
Special care should be taken when the left and right states of the Riemann problem are associated to different fluids (across the interface). In such conditions, the ALE property of the scheme is used to impose the interface velocity to be the velocity obtained from the Riemann solver \cite{Leduc2010}. This will therefore block the mass transfer across the interface. Following \cite{Guilcher2010} the equation for particles' volume evolution is modified as
\begin{equation}\label{omega_sph_ale}
\frac{\ud (\omega _i)}{\ud t} = \omega _i \sum _j \omega _j \left( {\bf v}_{E,ij} - {\bf v}_0 \left( {\bf x}_i\right) \right) B_{ij} \cdot  \nabla _i W_{ij},
\end{equation}
across the interface. The presented solution of the Riemann problem is based on the Godunov first order upwind method which assumes piecewise constant data. However, the accuracy of this approach is generally not sufficient due to the dissipative nature of low order schemes. The MUSCL (Monotone Upstream-centered Schemes for Conservation Laws) scheme is used to extend the accuracy of the proposed SPH formulation to second order (see \cite{Rafiee2012, Rafiee2012a} for complete details on the implementation of the MUSCL algorithm in SPH).

\subsection{Tensile Instability}

\textsc{Swegle} \emph{et al.} (1995) \cite{Swegle1995} have performed a one--dimensional von Neumann stability analysis of the SPH method and found that the method is unstable for particle $i$ if $\sum_j W''\bigl(r_{ij},h\bigr) T_i > 0$, where $W''\bigl(r_{ij},h\bigr)$ is the second derivative of the kernel and $T_i$ is the stress on particle $i$ which is negative under compression and positive under tension. This is the so-called \textquotedblleft Tensile instability \textquotedblright in the SPH literatures. Tensile instability results in the particles tendency to clump together. Robinson (2009) \cite{Robinson2009} showed that this behaviour of particles is directly related to a property of the SPH kernel. In the case of the cubic spline kernel, this is the location of the spline point. It was then concluded that the spline point must be set to the initial particle spacing in order to minimise particle clumping \cite{Robinson2009}. However, \textsc{Monaghan} (2000) \cite{Monaghan2000} revealed that the tensile instability can be alleviated using an artificial pressure in the momentum equation.

Nevertheless, none of the aforementioned techniques remove the tensile instability completely and it is only palliated by these corrections. Here, a new approach is proposed to remove the tensile instability. To do so, it is found that this instability occurs when the pressure becomes negative. Therefore, a constant background pressure was added to the equation of state in both liquid and gaseous phases. The equation of state then takes the form
\begin{equation}\label{eos_p0}
  p_i = \frac{{\rho_0}_i{c^2_0}_i}{\gamma_i}\left[\Bigl(\frac{\rho_i}{{\rho_0} _i}\Bigr)^{\gamma} - 1\right] + p_0
\end{equation}  
It should be noted that this correction is not applicable when the standard SPH formulation is used.

\subsection{Boundary Conditions}

\textsc{Vila} (1999) \cite{Vila1999} discussed the implementation of various boundary conditions in the SPH--ALE context. More recently, \cite{Marongiu2008, DeLeffe2009} proposed and implemented a novel boundary condition in the SPH formulations based on a reconstruction of the surface elements on the boundaries and on the solution to a partial Riemann problem \cite{Dubois2001} when a fluid particle interacts with a boundary surface element. Although this novel boundary condition looks promising however the implementation to multi-fluid flows and during strong impacts (with gas entrainment) is not straightforward.

Here, we model boundaries with the ghost particle approach. The ghost particle approach is based on mirroring the fluid particles on the other side of the boundary. The ghost boundary method provides a very accurate and stable boundary condition and has been used extensively by SPH practitioners \cite{Colagrossi2003, Marrone2011}.

\subsection{Time Stepping}

The SPH--ALE sets of equations (\ref{sph_ale}) can be marched in time using any stable time integrating algorithm for ordinary differential equations. Here, a second order symplectic time integration scheme is used to calculate the evolution of the SPH--ALE equations in time. \textsc{Guilcher} \emph{et al.} (2010) \cite{Guilcher2010} suggested the use of classical $4$\up{th} order Runge--Kutta or $3$\up{rd} order Strong Stability Preserving Rung--Kutta schemes. However, higher order schemes significantly increase the computational cost which does not necessarily improve the accuracy of the results.

The time step is in general restricted by a CFL condition on acoustic waves.

\section{Results and Discussions}\label{sec:results}

Here, two different problems are simulated using the proposed SPH scheme. The first test case compares the results of the SPH scheme in capturing a sharp impact pressure with an incompressible Level Set solver (for a description of the Level Set solver see \cite{Rafiee2013}). In the second test case the wave impact on a rigid wall with a large entrained air pocket is simulated. The SPH simulation is initialised by the output of a fully non-linear potential flow solver \cite{Scolan2010}.

\subsection{The Liquid Patch impact Test Case}

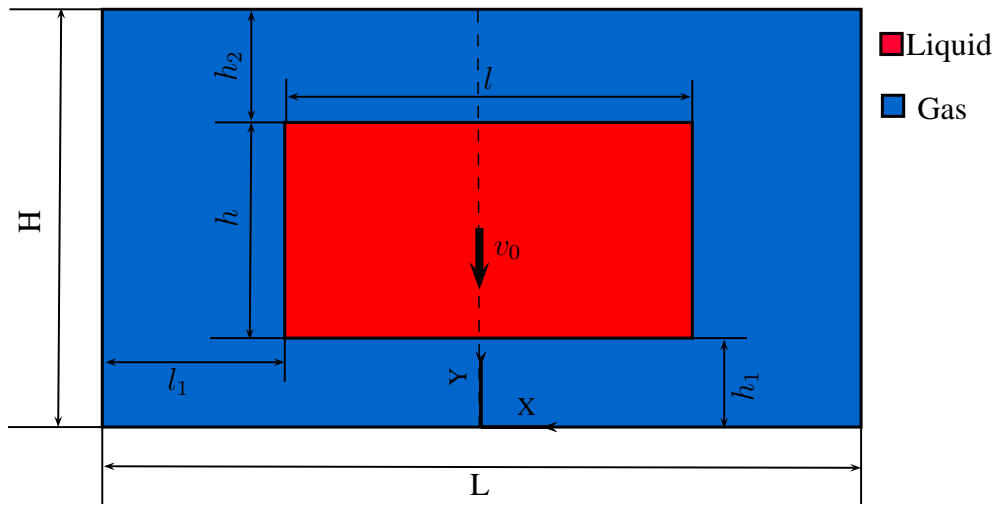
\begin{figure}
  \centering
\scalebox{1} 
{
\begin{pspicture}(0,-3.303)(12.880938,3.303)
\definecolor{color6826b}{rgb}{0.0,0.4,0.8}
\definecolor{color6878b}{rgb}{1.0,0.0,0.2}
\psframe[linewidth=0.04,dimen=outer,fillstyle=solid,fillcolor=color6826b](11.24,3.29)(1.22,-2.29)
\psframe[linewidth=0.04,dimen=outer,fillstyle=solid,fillcolor=red](9.02,1.79)(3.62,-1.11)
\psline[linewidth=0.025999999cm](1.24,-2.27)(1.24,-3.29)
\psline[linewidth=0.025999999cm](11.22,-2.27)(11.22,-3.29)
\psline[linewidth=0.025999999cm,arrowsize=0.05291667cm 2.0,arrowlength=1.4,arrowinset=0.4]{<->}(1.26,-2.79)(11.22,-2.81)
\psline[linewidth=0.025999999cm](1.24,-2.27)(0.0,-2.27)
\psline[linewidth=0.025999999cm](1.26,3.27)(0.02,3.27)
\psline[linewidth=0.025999999cm,arrowsize=0.05291667cm 2.0,arrowlength=1.4,arrowinset=0.4]{<->}(0.66,-2.27)(0.7,3.29)
\psline[linewidth=0.025999999cm](3.64,-1.09)(2.66,-1.09)
\psline[linewidth=0.025999999cm](3.7,1.77)(2.72,1.77)
\psline[linewidth=0.025999999cm,arrowsize=0.05291667cm 2.0,arrowlength=1.4,arrowinset=0.4]{<->}(3.18,-1.11)(3.2,1.77)
\psline[linewidth=0.025999999cm,arrowsize=0.05291667cm 2.0,arrowlength=1.4,arrowinset=0.4]{<->}(3.2,1.79)(3.2,3.25)
\psline[linewidth=0.02cm,linestyle=dashed,dash=0.16cm 0.16cm](6.2,-2.27)(6.18,3.25)
\rput(6.22,-2.27){\psaxes[linewidth=0.04,ticksize=0.10583333cm,showorigin=false]{-<}(0,0)(0,0)(1,1)}
\usefont{T1}{ptm}{m}{n}
\rput(6.211406,-3.025){$\textsc{L}$}
\psline[linewidth=0.025999999cm](8.96,-1.09)(9.72,-1.09)
\psline[linewidth=0.025999999cm,arrowsize=0.05291667cm 2.0,arrowlength=1.4,arrowinset=0.4]{<->}(9.42,-1.11)(9.42,-2.29)
\psline[linewidth=0.025999999cm](3.64,-1.09)(3.64,-1.67)
\psline[linewidth=0.025999999cm,arrowsize=0.05291667cm 2.0,arrowlength=1.4,arrowinset=0.4]{<->}(1.28,-1.41)(3.64,-1.41)
\psline[linewidth=0.106000006cm,arrowsize=0.05291667cm 2.0,arrowlength=1.4,arrowinset=0.4]{->}(6.2,0.37)(6.2,-0.45)
\usefont{T1}{ptm}{m}{n}
\rput(6.8214064,-2.005){$\textsc{x}$}
\usefont{T1}{ptm}{m}{n}
\rput{-270.0}(4.3501563,-7.529219){\rput(5.920625,-1.5748438){$\textsc{y}$}}
\usefont{T1}{ptm}{m}{n}
\rput(2.2414062,-1.665){$l_1$}
\usefont{T1}{ptm}{m}{n}
\rput{-270.0}(8.010156,-11.409219){\rput(9.690625,-1.6848438){$h_1$}}
\usefont{T1}{ptm}{m}{n}
\rput{-270.0}(3.385,-2.3440626){\rput(2.8454688,0.51){$h$}}
\usefont{T1}{ptm}{m}{n}
\rput{-270.0}(5.450156,-0.36921883){\rput(2.890625,2.5551562){$h_2$}}
\usefont{T1}{ptm}{m}{n}
\rput{-270.0}(0.77015626,0.17078125){\rput(0.280625,0.48515624){$\textsc{H}$}}
\psline[linewidth=0.025999999cm](3.66,2.35)(3.66,1.77)
\psline[linewidth=0.025999999cm](9.0,2.33)(9.0,1.75)
\psline[linewidth=0.025999999cm,arrowsize=0.05291667cm 2.0,arrowlength=1.4,arrowinset=0.4]{<->}(3.68,2.11)(8.98,2.11)
\usefont{T1}{ptm}{m}{n}
\rput(6.3214064,2.355){$l$}
\psframe[linewidth=0.04,dimen=outer,fillstyle=solid,fillcolor=color6878b](11.78,2.97)(11.46,2.65)
\psframe[linewidth=0.04,dimen=outer,fillstyle=solid,fillcolor=color6826b](11.8,2.15)(11.48,1.83)
\usefont{T1}{ptm}{m}{n}
\rput(12.287344,1.955){Gas}
\usefont{T1}{ptm}{m}{n}
\rput(12.376875,2.755){Liquid}
\usefont{T1}{ptm}{m}{n}
\rput(6.5714064,0.075){$v_0$}
\end{pspicture} 
}
  \caption{\small\em Liquid Patch impact problem.}
  \label{impact_dimensions}
\end{figure}

\begin{figure}
\centering
\input{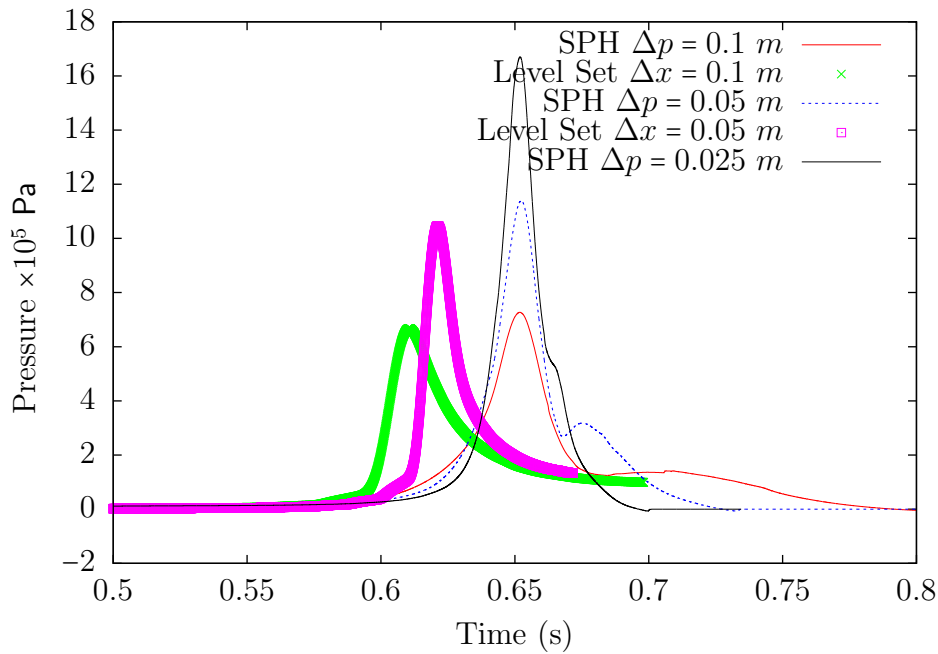}      
\caption{\small\em Comparison of the time variation of the impact pressure at the centre of the bottom wall between the SPH and the Level Set simulations at different resolutions.}
\label{impactpress}
\end{figure}

The problem studied here is the impact of a liquid patch on a rigid horizontal wall \cite{Braeunig2009}. The initial shape of the liquid patch is rectangular and is at rest in an atmosphere of a gaseous phase at time $t=0$. The liquid patch then falls freely under gravity. The dimensions of the problem are shown in Figure~\ref{impact_dimensions} and dimensions are given in Table~\ref{tab:patch}. The simulations were performed for water (heavy fluid) and air (the light fluid). The physical properties of the fluids are given in Table~\ref{characteristics}. Figure~\ref{impactpress} compares the impact pressure obtained with the SPH simulations and with the Level Set simulations at different resolutions. The impact pressure was measured at the centre of the bottom wall.

\begin{table}
\centering
\begingroup\setlength{\fboxsep}{0pt}
\colorbox{lightgray}{
\begin{tabular}{c|c}
   \textit{Dimension} & $\mathsf{m}$ \\ \hline\hline
   H & 15 \\ \hline
   $h$ & 8 \\ \hline
   $h_1$ & 2 \\ \hline
   $h_2$ & 5 \\ \hline
   L & 20 \\ \hline
   $l$ & 10 \\ \hline
   $l_1$ & 5 \\
\end{tabular}}\endgroup
\caption{\small\em Parameters of the Liquid Patch impact problem.}
\label{tab:patch}
\end{table}

\begin{table}
\centering
\begingroup\setlength{\fboxsep}{0pt}
\colorbox{lightgray}{
\begin{tabular}{c|c|c}
  & \multicolumn{1}{|c|}{\textit{Water}} & \multicolumn{1}{|c}{\textit{Air}} \\
\hline\hline 
Density $(\rho)~\mathsf{kg}\,\mathsf{m}^{-3}$ & 1000 & 1.2 \\
\hline Sound Speed $(c)~\mathsf{m}\,\mathsf{s}^{-1}$ & 1500 & 342 \\
\hline Isentropic Exponent $(\gamma)$ & 7 & 4 \\
\end{tabular}
}\endgroup
\caption{\small\em Properties of fluids for the Liquid Patch impact problem.}
\label{characteristics}
\end{table}

It can clearly be seen that both methods agree very well in terms of the magnitude of the impact pressure at the corresponding resolution. However, the temporal location of the impact differs between SPH and Level Set. This difference is also visible between Level Set results at different resolutions. This is due to the fact that the interface is sharp in the SPH method whereas it has a thickness (of $\sim 2 \Delta x$) in the Level Set scheme. Therefore, the interface in the Level Set simulations is thicker in coarser resolutions and hence the impact occurs sooner. It is worth noting that although the Level Set solver is incompressible however since the interface is not sharp and has a thickness, the scheme can be considered compressible across the interface.

\subsection{Wave Impact on a Rigid Wall}

In the previous test case it was shown that the proposed two-phase compressible SPH scheme is capable of modelling violent impacts accurately. However, in case of a wave impact on a structure, the generation and propagation of the wave can take a long time with the SPH method. Therefore, like in \cite{Guilcher2012}, a potential flow solver (FSID code; incompressible, inviscid and single fluid with free-surface solver) is used to generate and propagate the wave up to the impact point. The output of the potential flow solver is then used to initialise the SPH particles. Although the FSID code is very efficient for wave generation and propagation, it cannot model the compressibility effects during the impact and it is not able to compute the solution when the wave crest approaches (hits) the wall.

\begin{figure}
\centering
\subfigure[]{\includegraphics[width=0.49\textwidth]{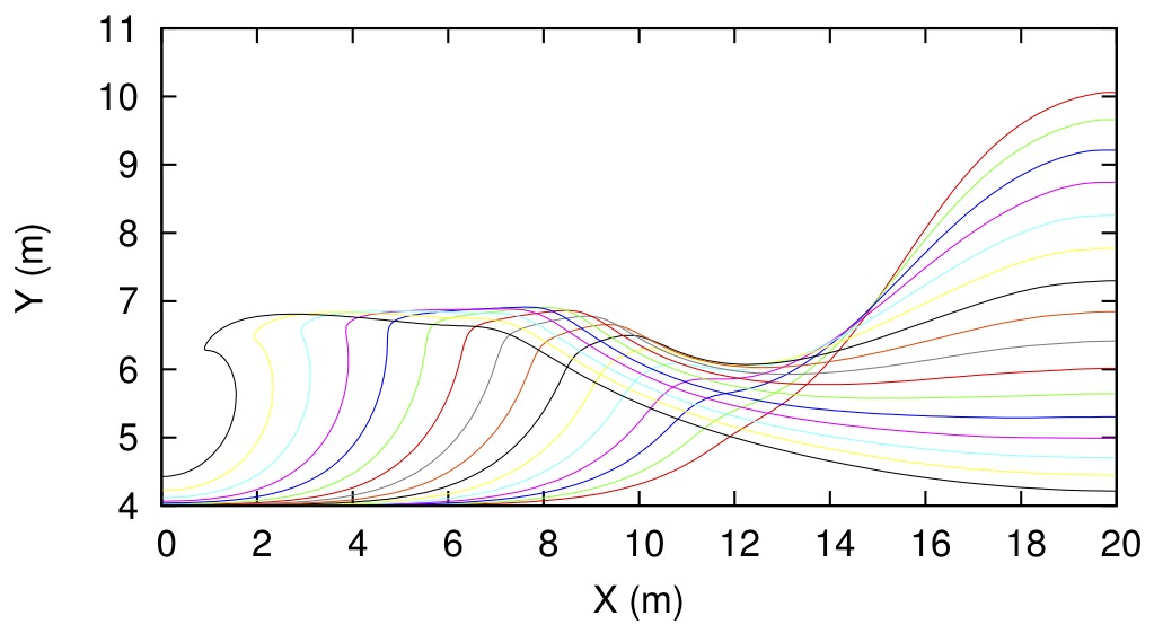}}
\subfigure[]{\includegraphics[width=0.49\textwidth]{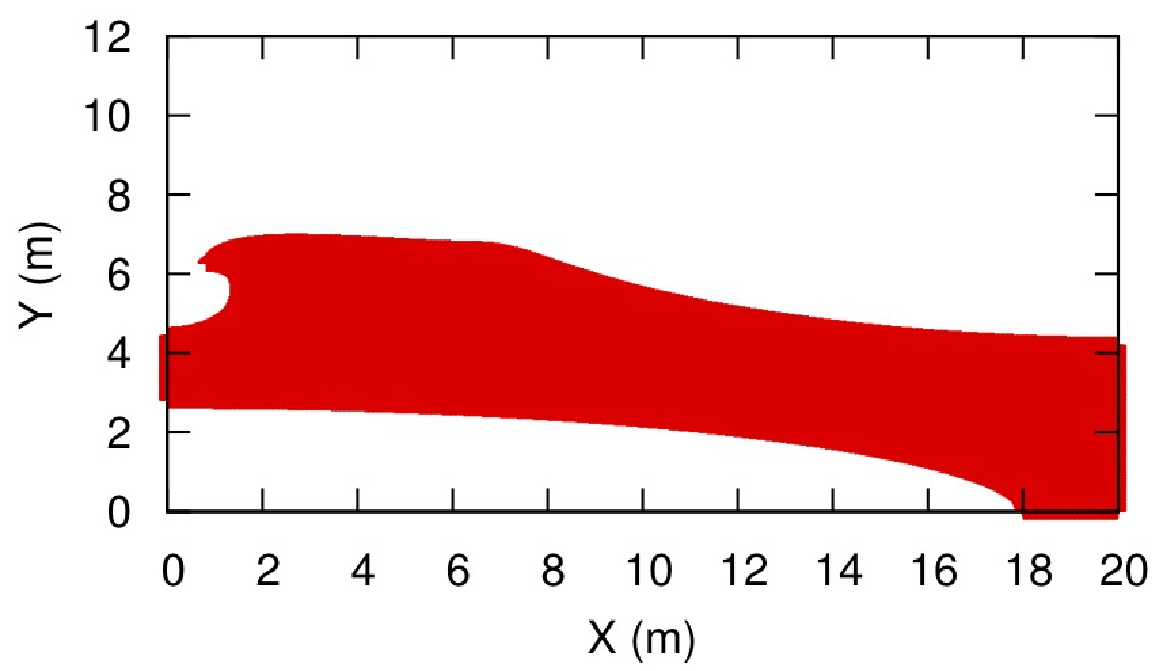}}
\caption{\small\em (a): Evolution of the desired wave with FSID. (b): Final shape of the wave used for initialising the SPH particles. }
\label{fsid_out}
\end{figure}

In order to initialise the SPH particles, a bilinear interpolation is used to map the interface profile, velocity and pressure fields from a fixed grid (provided by FSID) to the initial set-up of the particles. Here, only the particles in the liquid phase are initialised and the gas particles are at rest (zero velocity and pressure equal to the background pressure). Figure~\ref{fsid_out} shows the evolution and the final shape of the wave generated with the FSID code.

\begin{figure}
  \centering
  \subfigure[]{\includegraphics[width=0.49\textwidth]{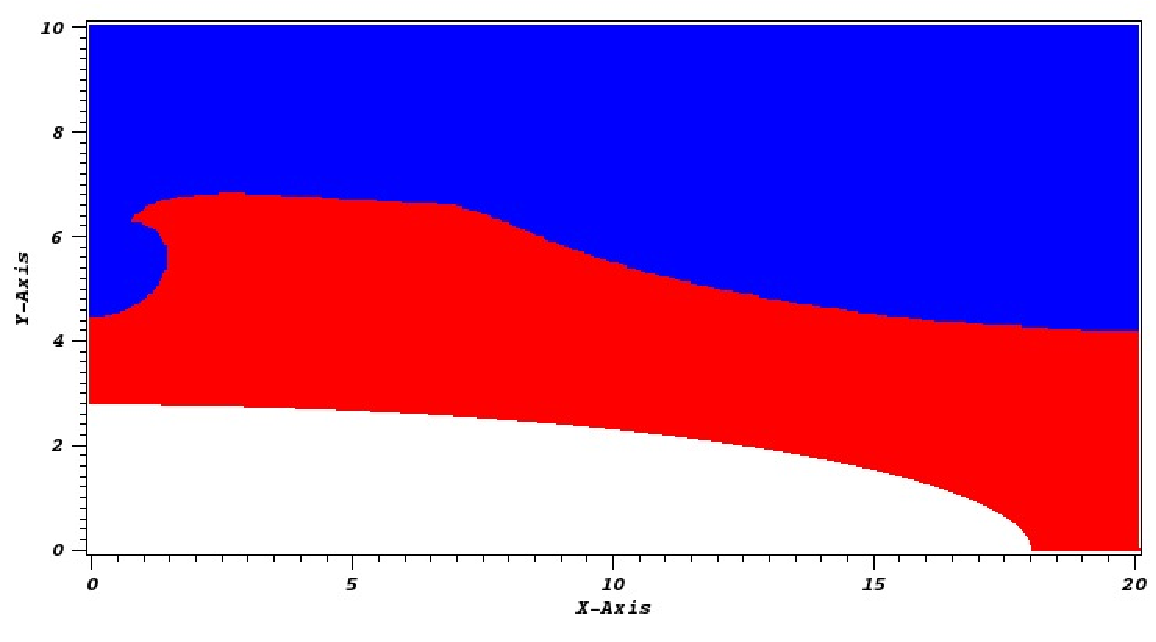}}
  \subfigure[]{\includegraphics[width=0.49\textwidth]{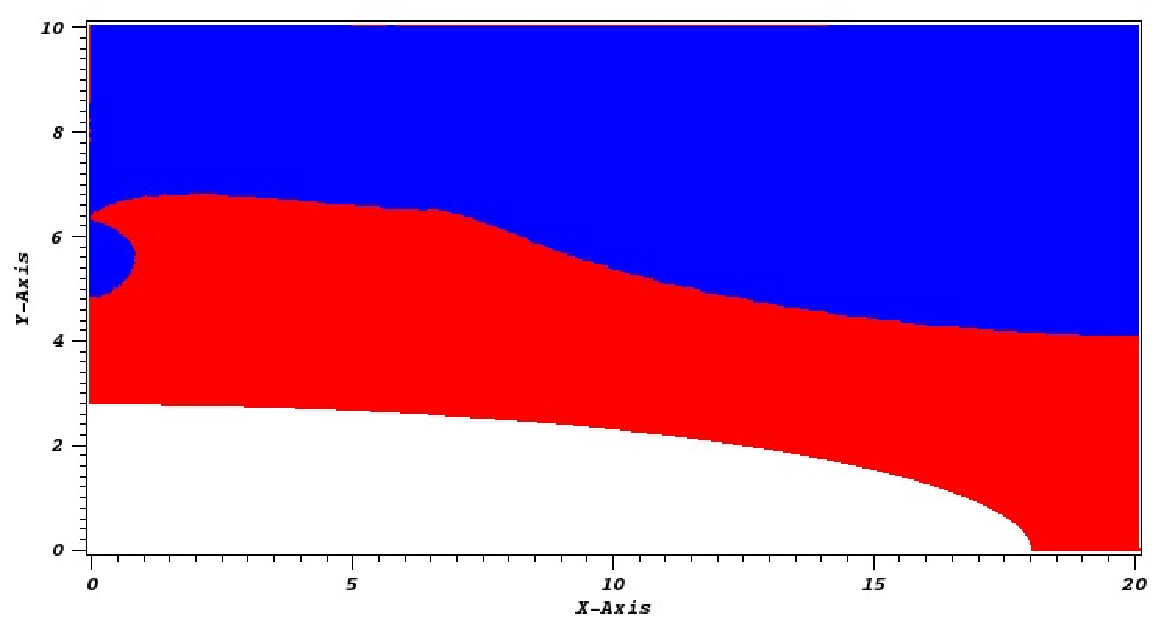}}
  \subfigure[]{\includegraphics[width=0.49\textwidth]{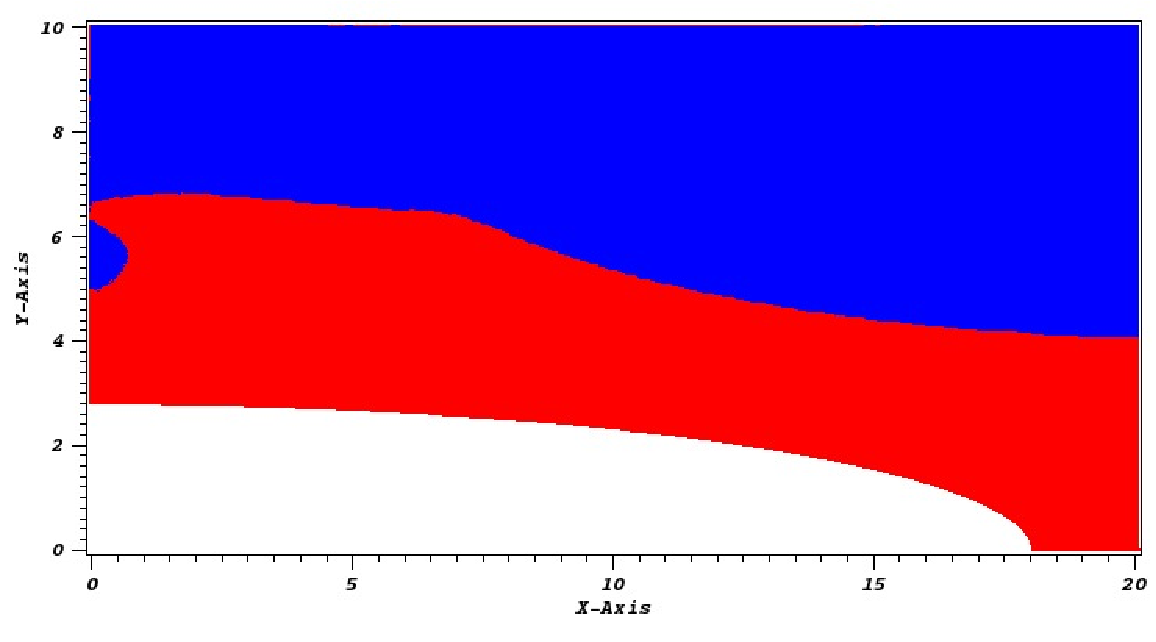}}
  \subfigure[]{\includegraphics[width=0.49\textwidth]{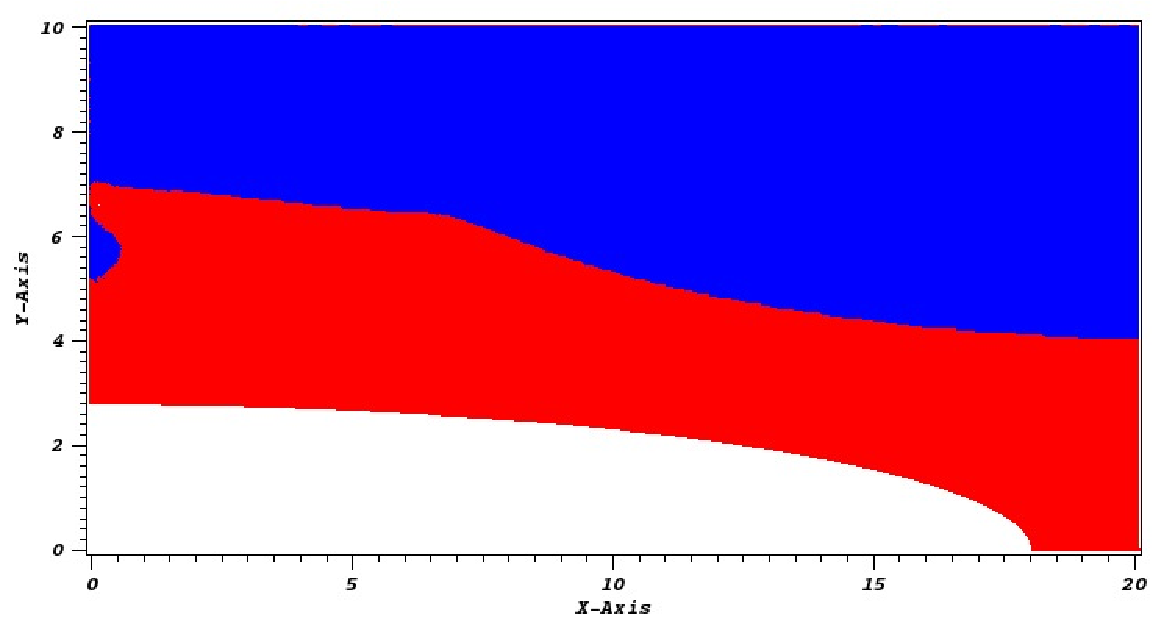}}
  \subfigure[]{\includegraphics[width=0.49\textwidth]{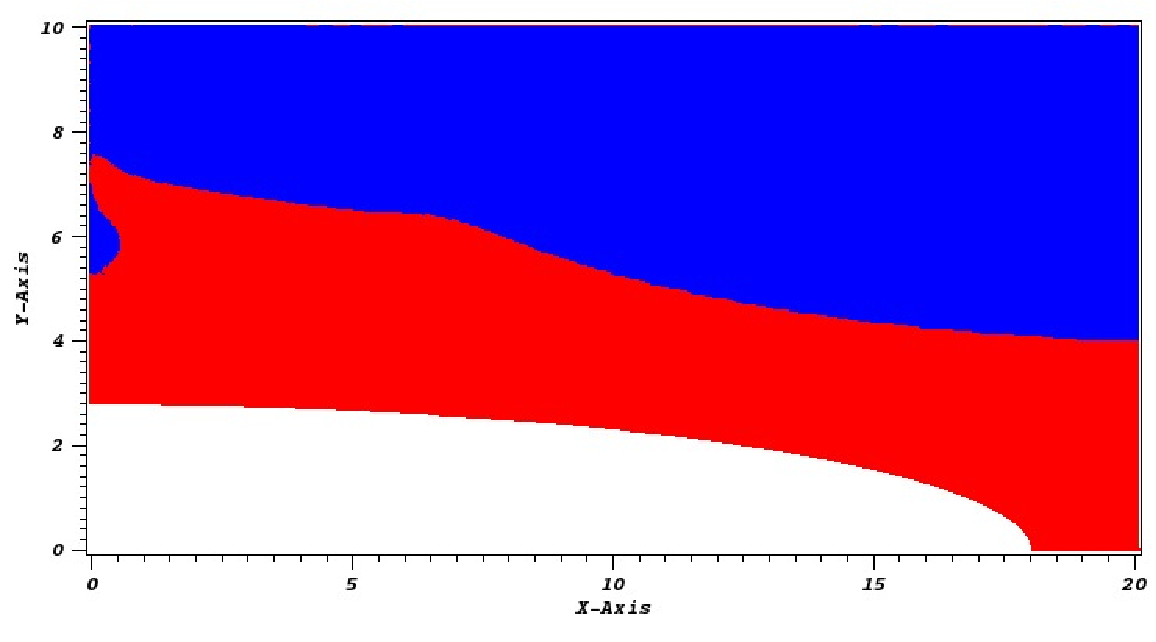}}
  \subfigure[]{\includegraphics[width=0.49\textwidth]{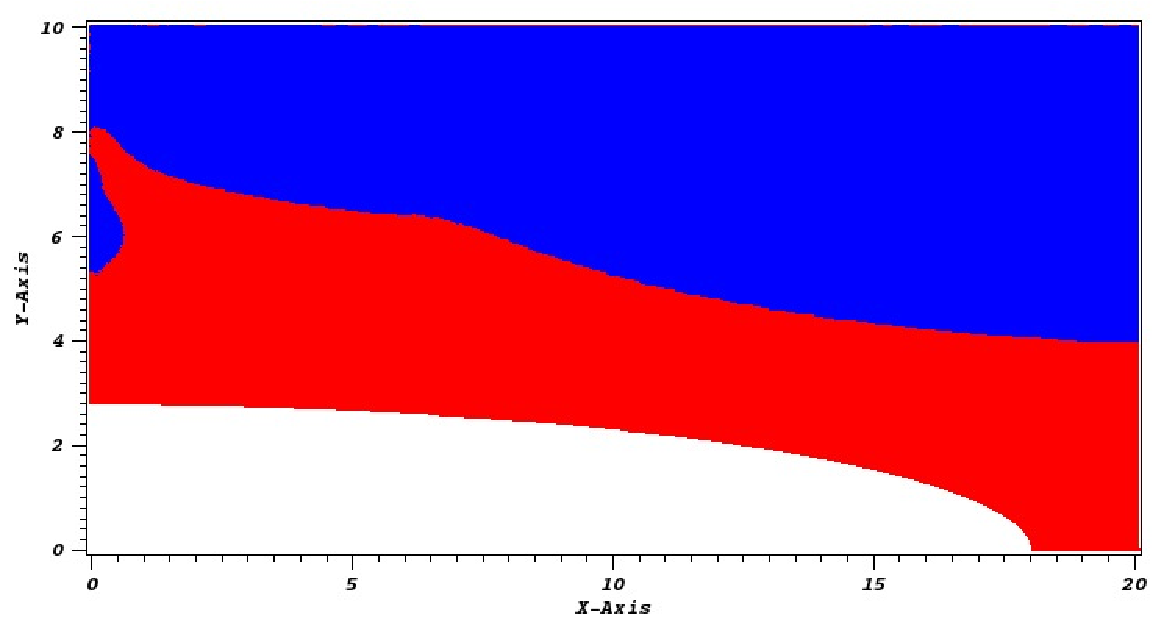}}
  \caption{\small\em Snapshots of the SPH simulations at various times. The initial spacing of the SPH particles was set to $0.033~\mathsf{m}$. The particles are coloured by their density; red and blue represent heavy fluid with $\rho = 1000$ $\mathsf{kg}\, \mathsf{m^{-3}}$ and light fluid with $\rho = 1.2$ $\mathsf{kg}\, \mathsf{m^{-3}}$, respectively. Note the sharp interface between the liquid and gaseous phases.}
  \label{sph_fsid}
\end{figure}

\begin{figure}
  \centering
  \subfigure[]{\includegraphics[width=0.32\textwidth]{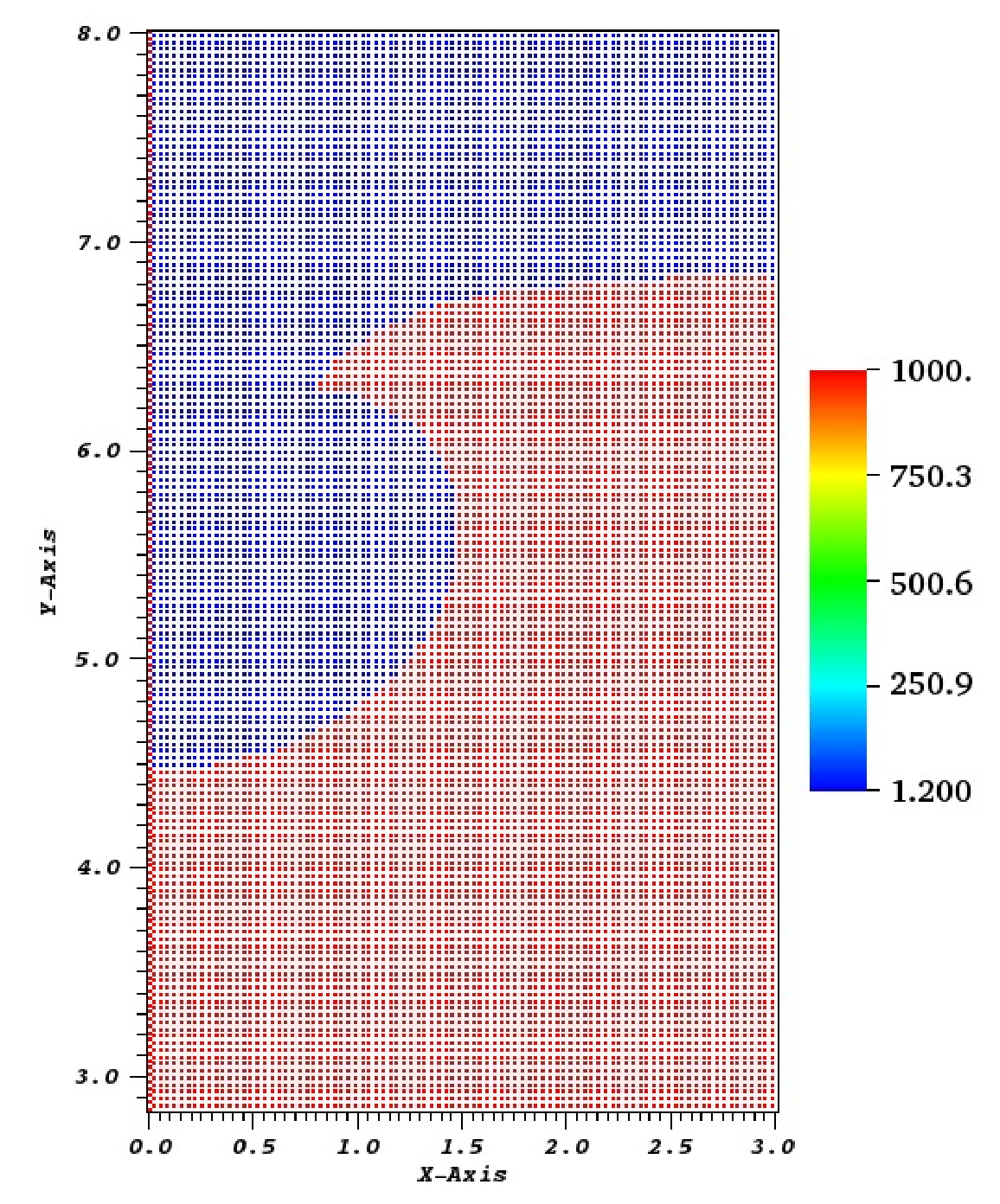}}
  \subfigure[]{\includegraphics[width=0.32\textwidth]{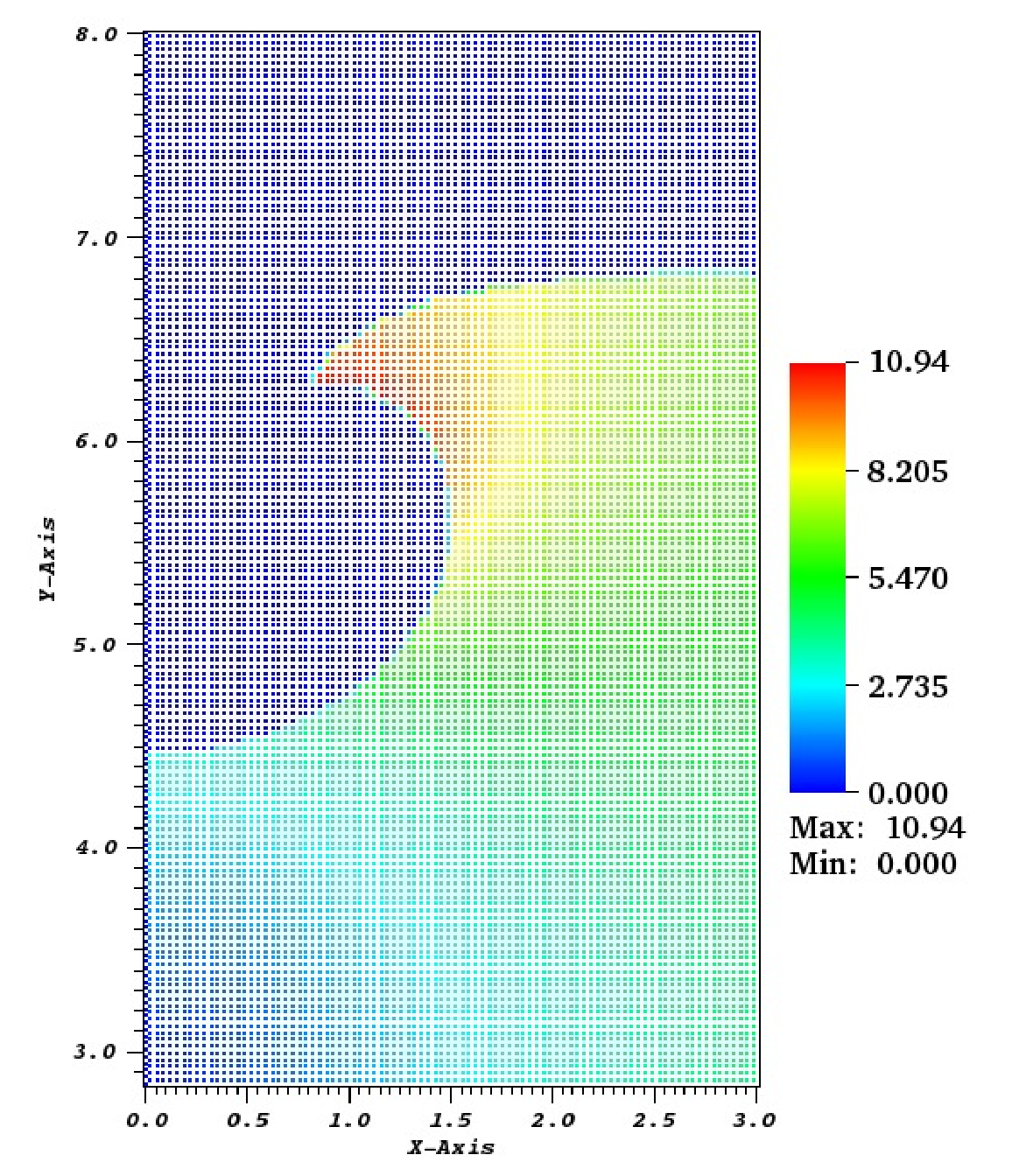}}
  \subfigure[]{\includegraphics[width=0.32\textwidth]{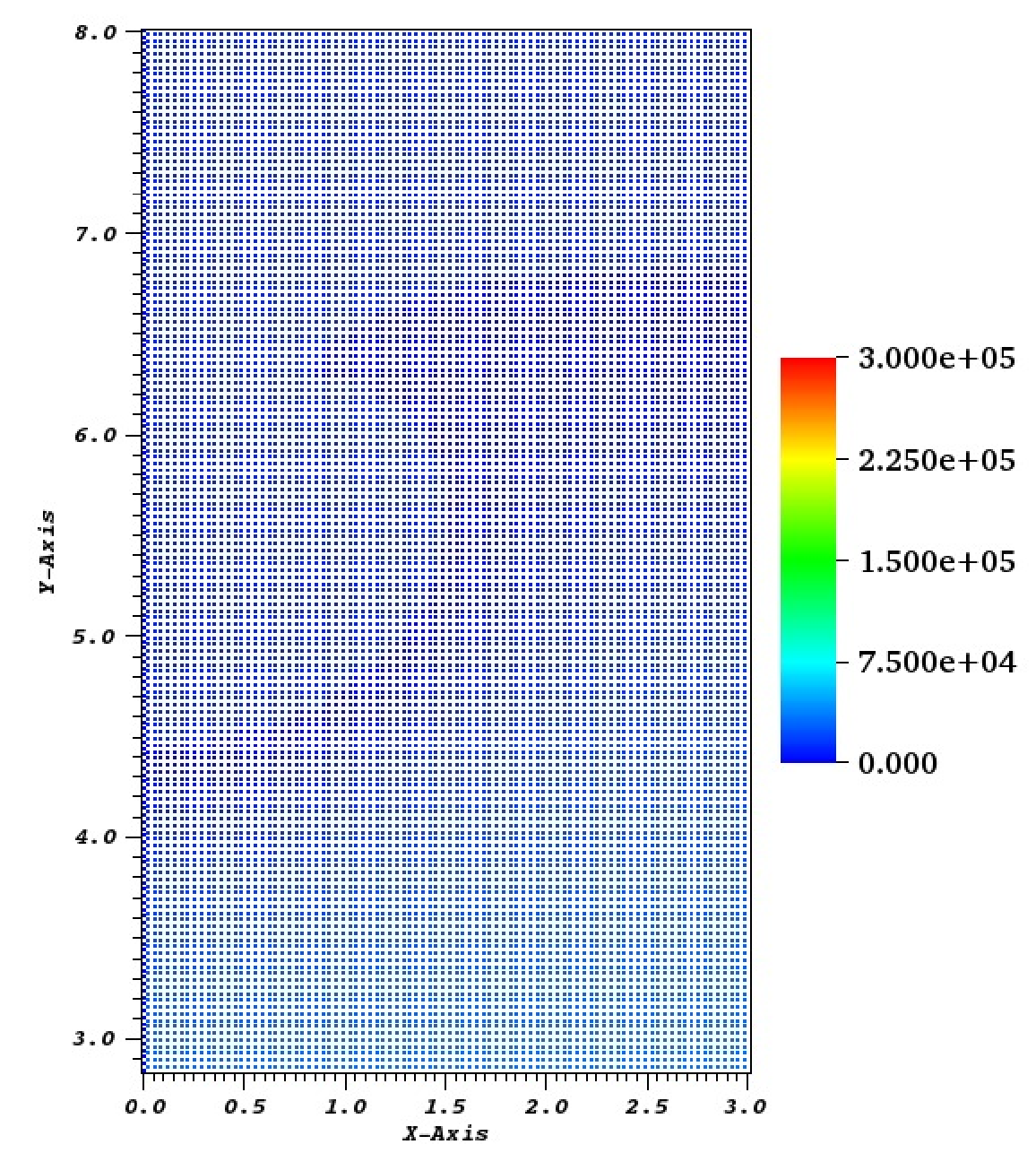}}
  \subfigure[]{\includegraphics[width=0.32\textwidth]{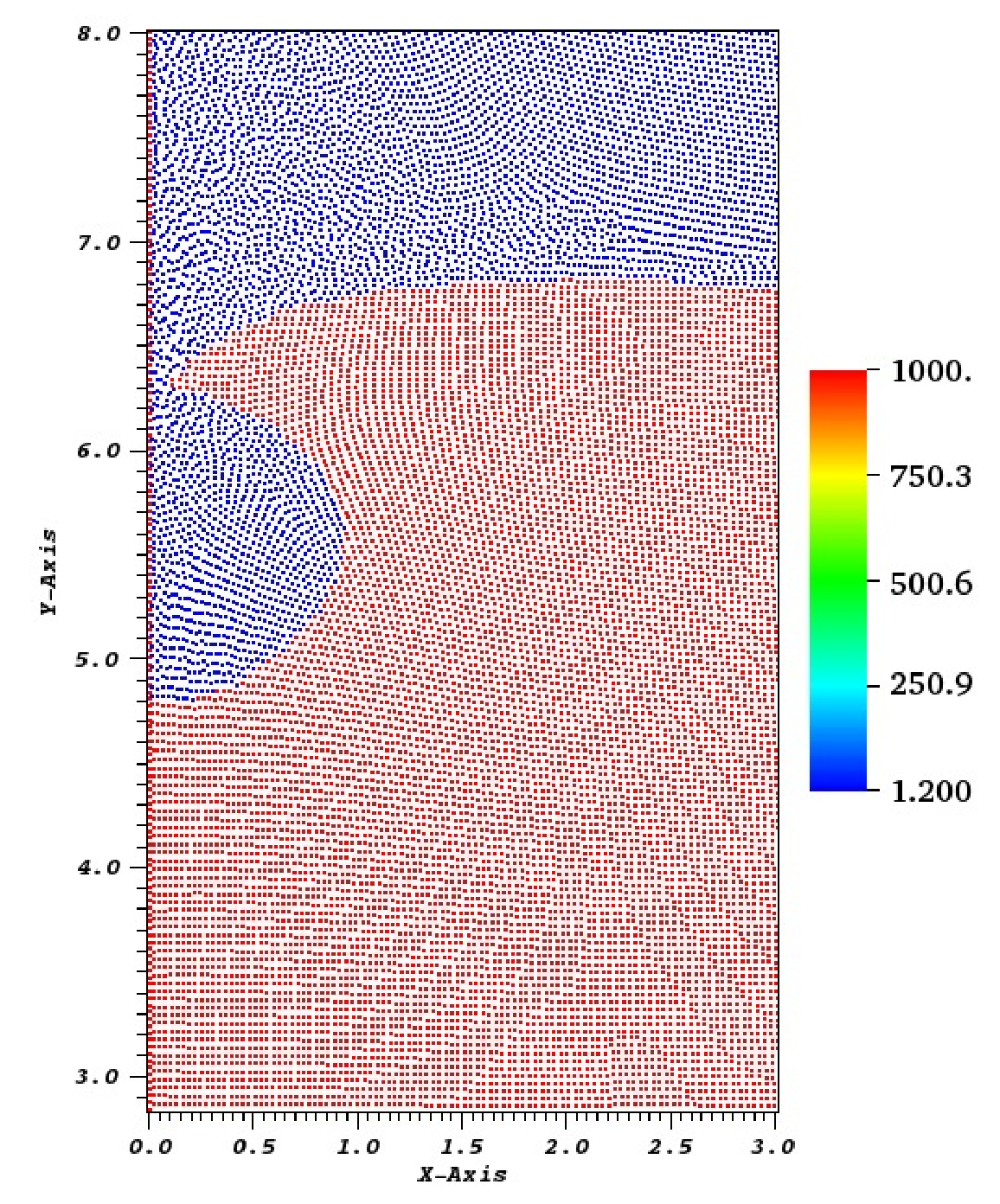}}
  \subfigure[]{\includegraphics[width=0.32\textwidth]{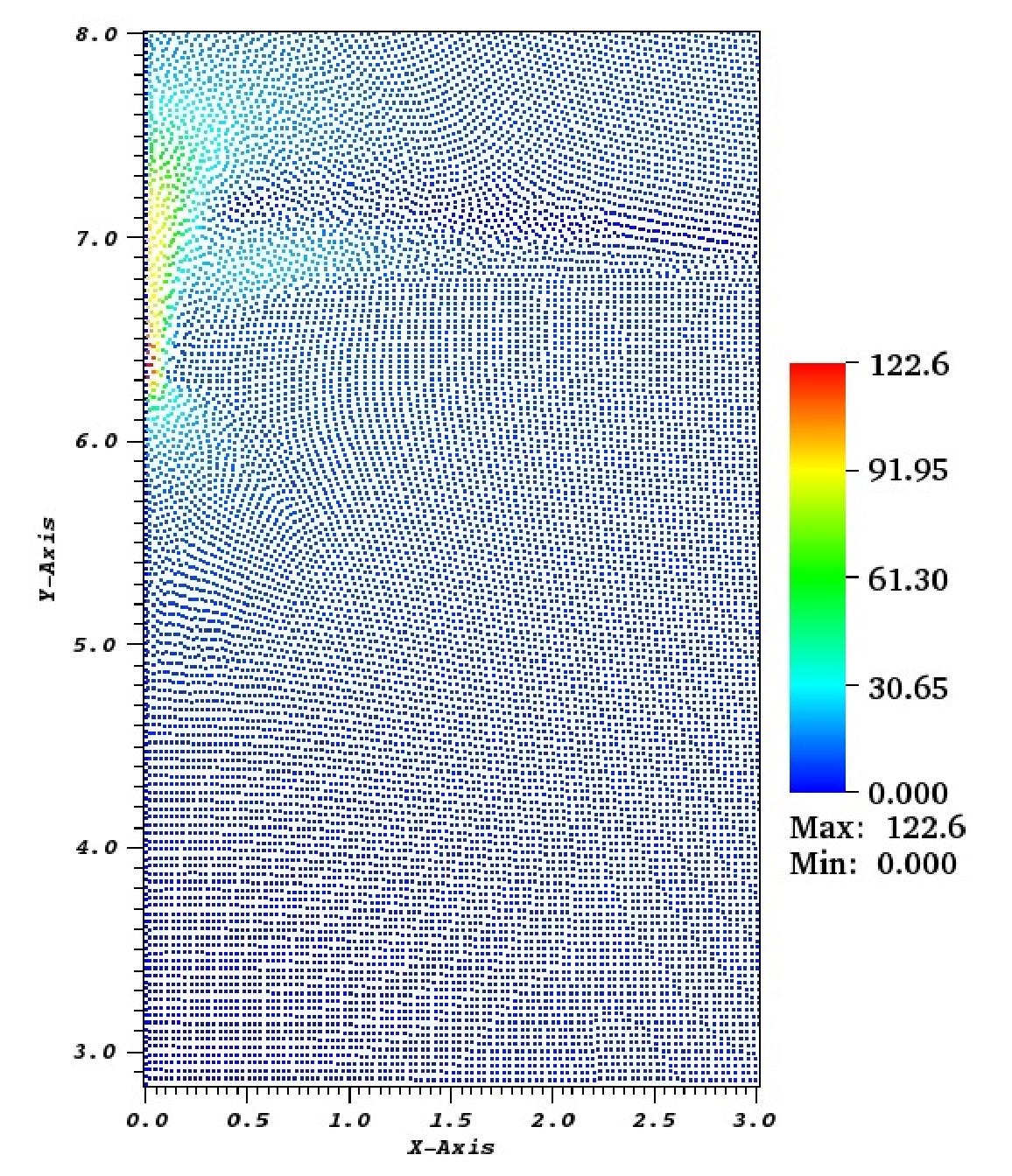}}
  \subfigure[]{\includegraphics[width=0.32\textwidth]{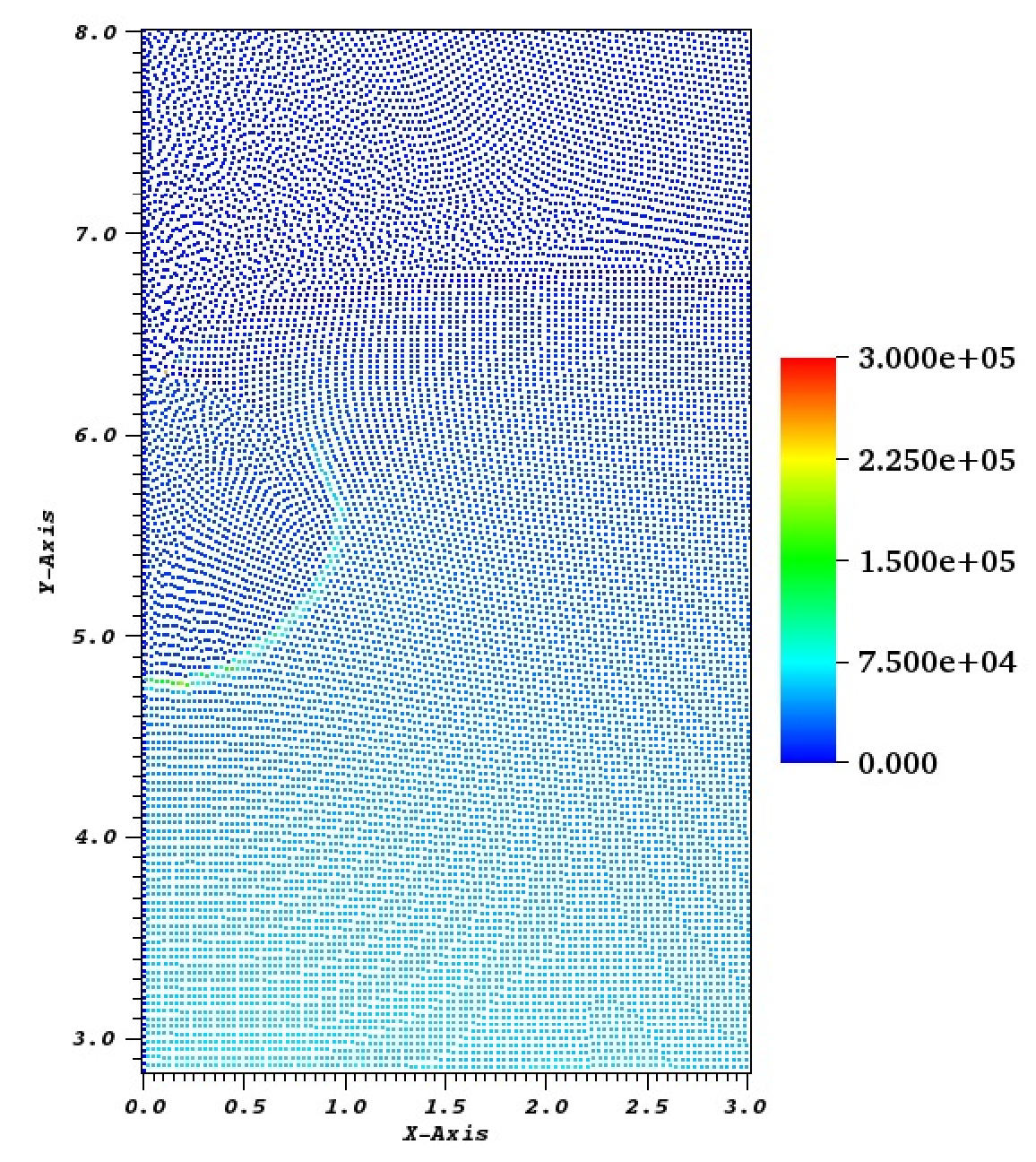}}
  \subfigure[]{\includegraphics[width=0.32\textwidth]{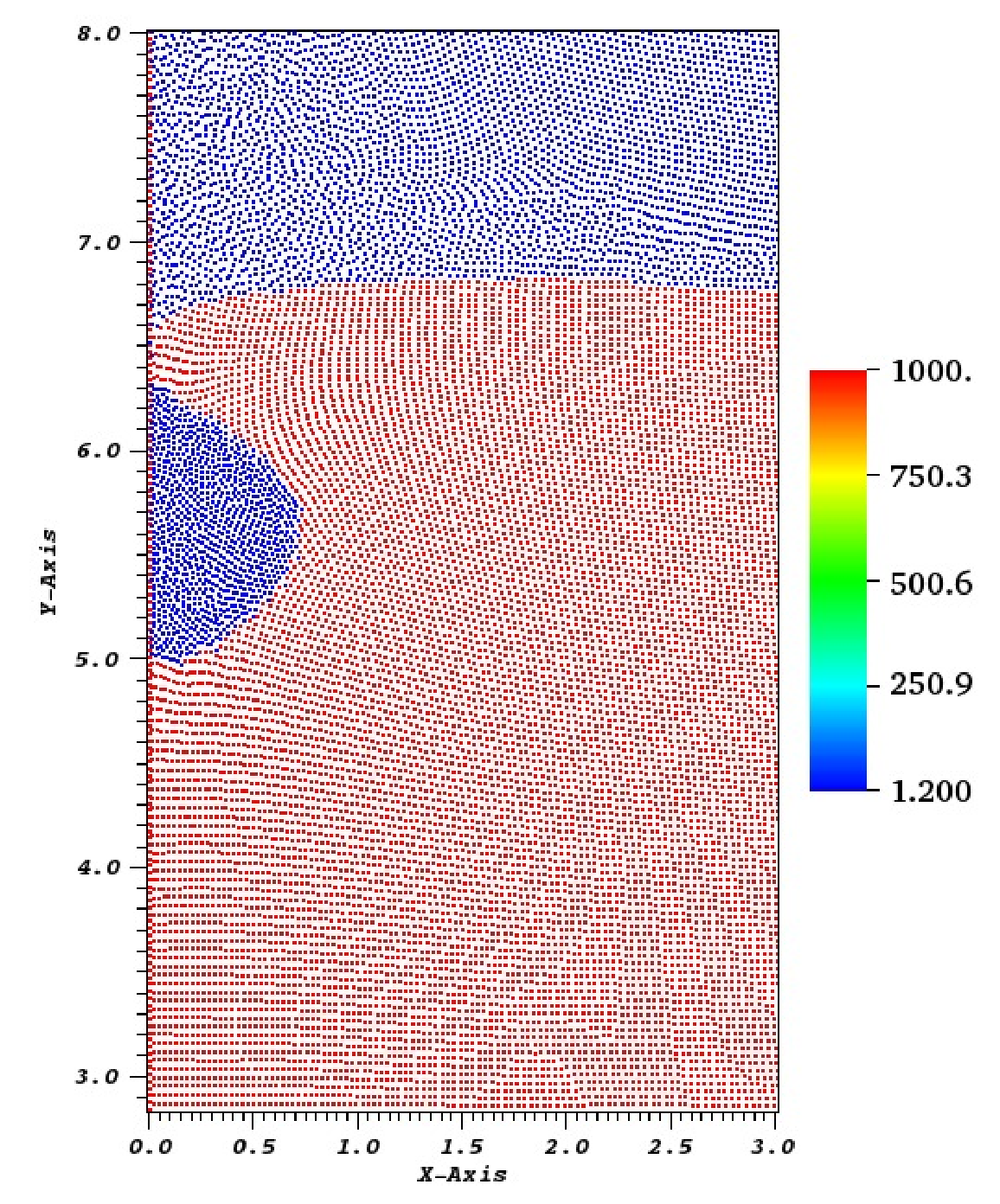}}
  \subfigure[]{\includegraphics[width=0.32\textwidth]{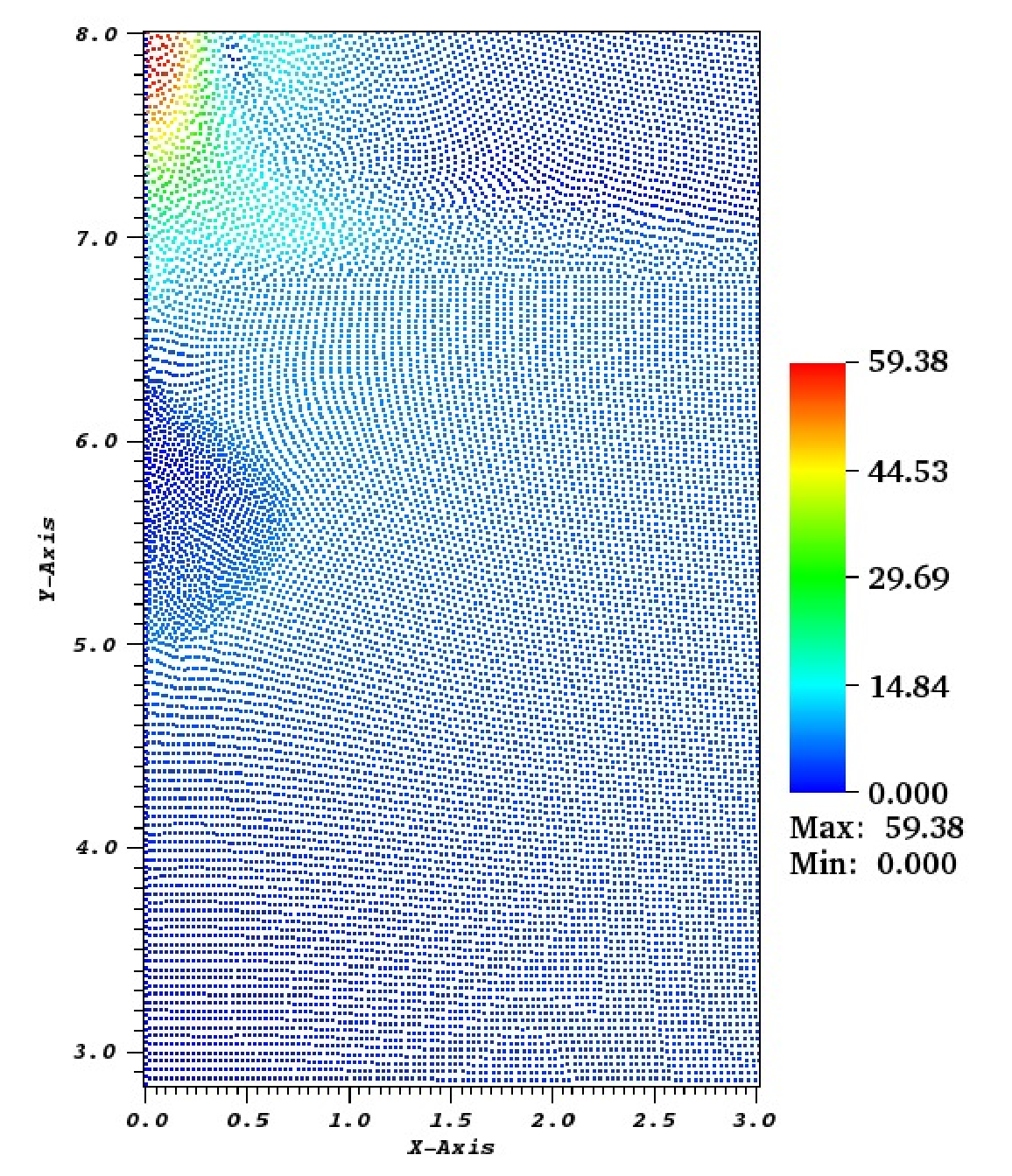}}
  \subfigure[]{\includegraphics[width=0.32\textwidth]{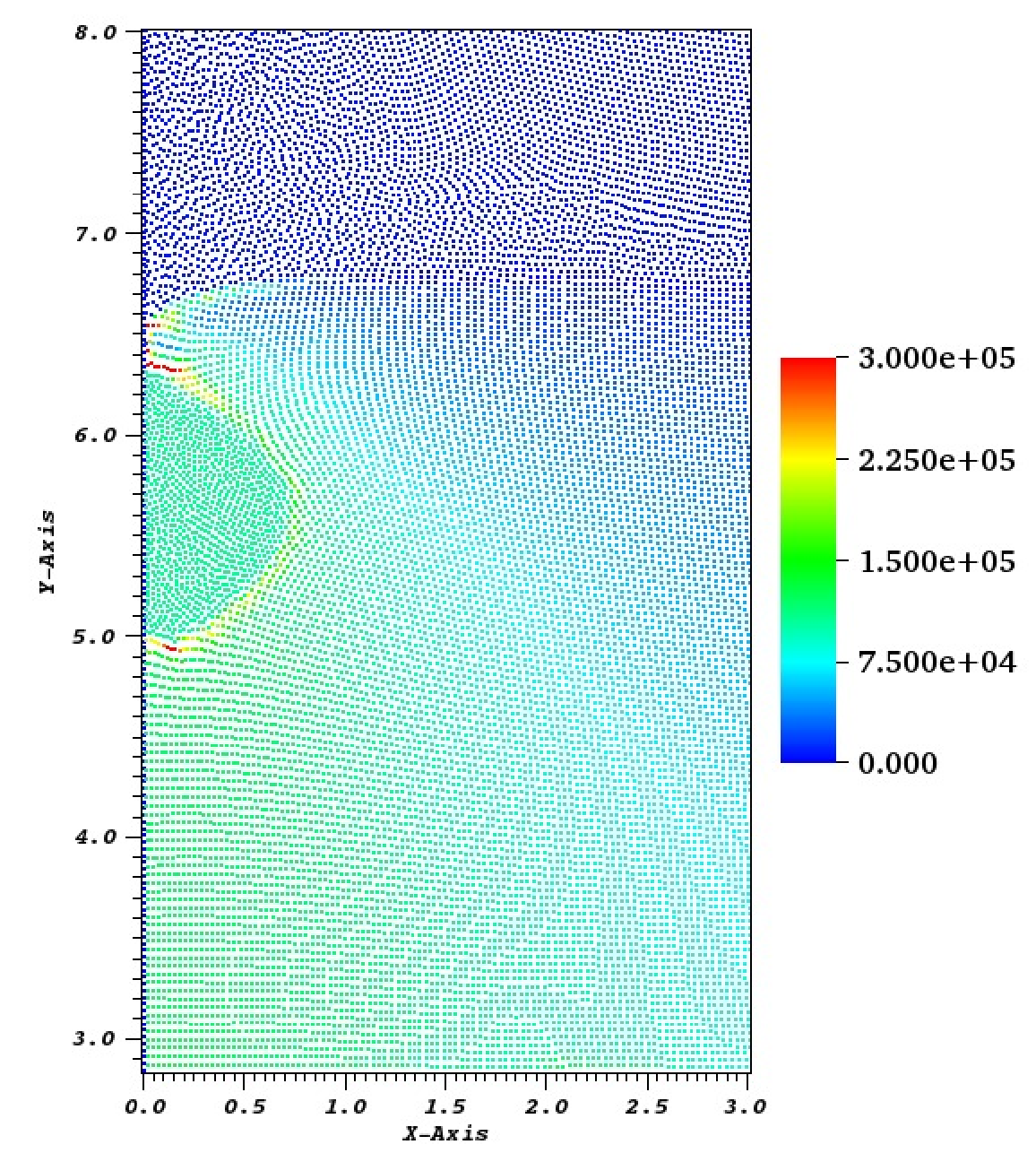}}
  \caption{\small\em Snapshots of the SPH simulations at various times. Particles are coloured by their density (left), velocity magnitude (middle) and pressure (right). Time increases from top to the bottom. Continued on the next page.}
  \label{sph_fsid_zoom}
\end{figure}

\begin{figure}
  \centering
  \subfigure[]{\includegraphics[width=0.32\textwidth]{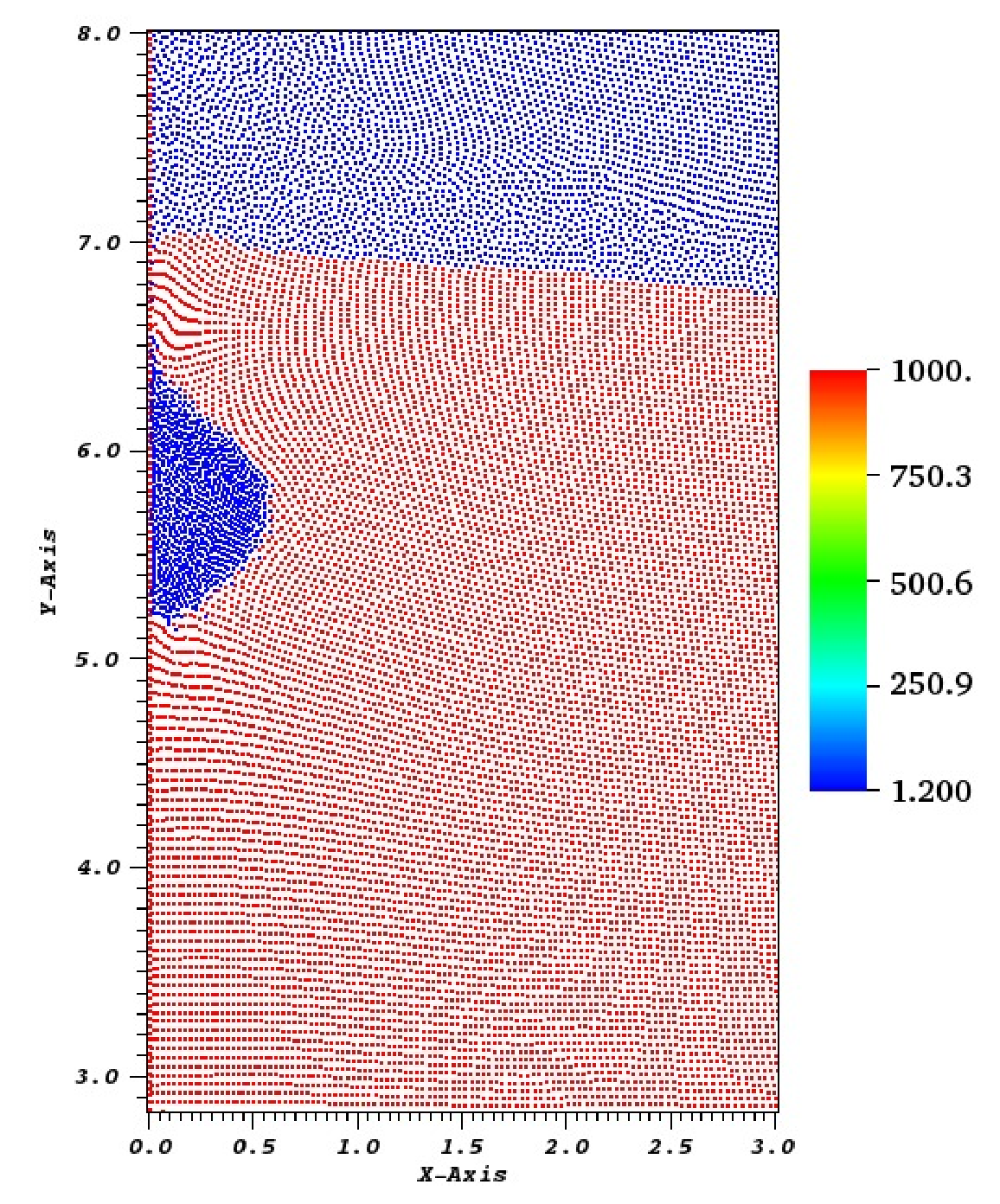}}
  \subfigure[]{\includegraphics[width=0.32\textwidth]{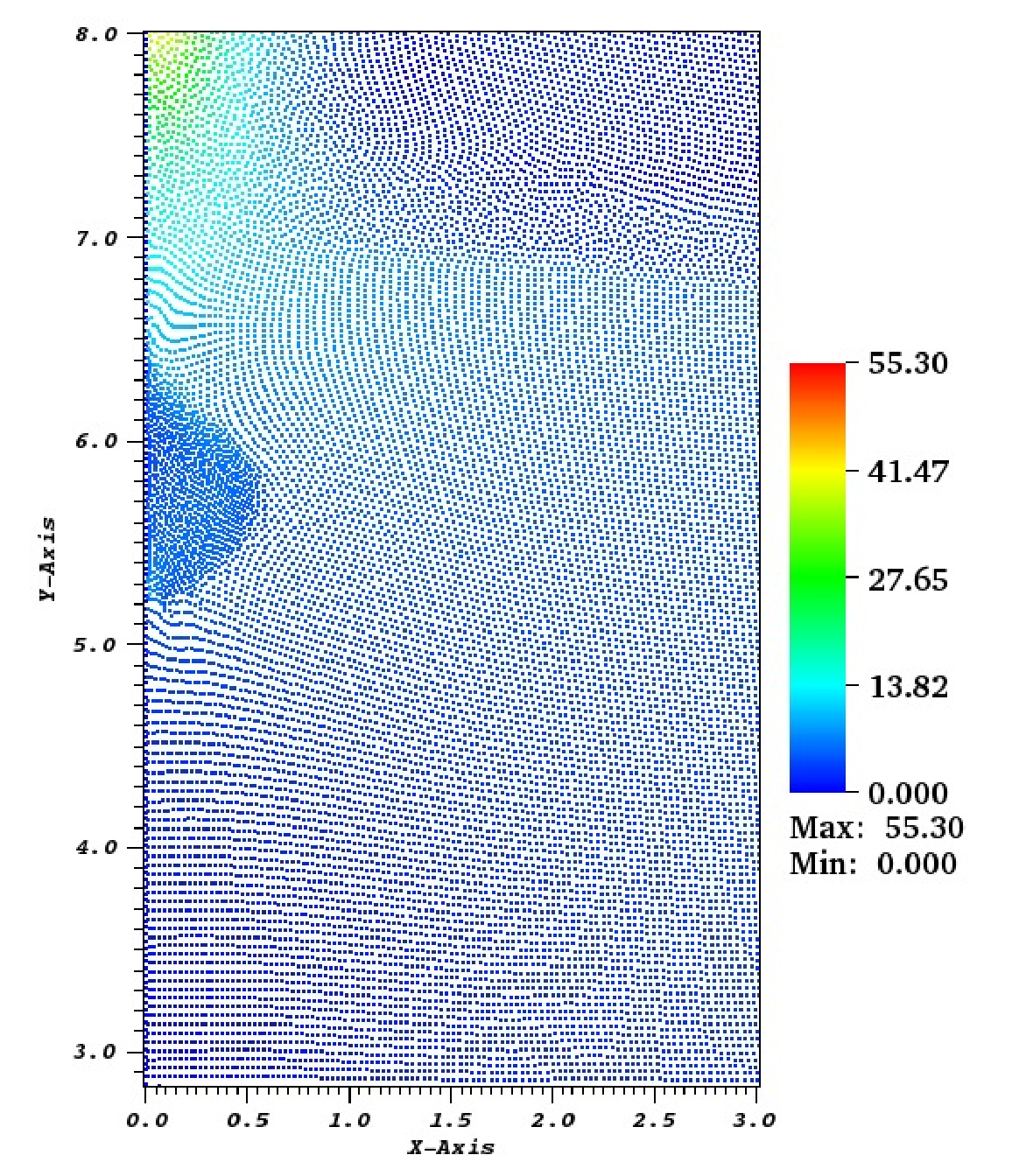}}
  \subfigure[]{\includegraphics[width=0.32\textwidth]{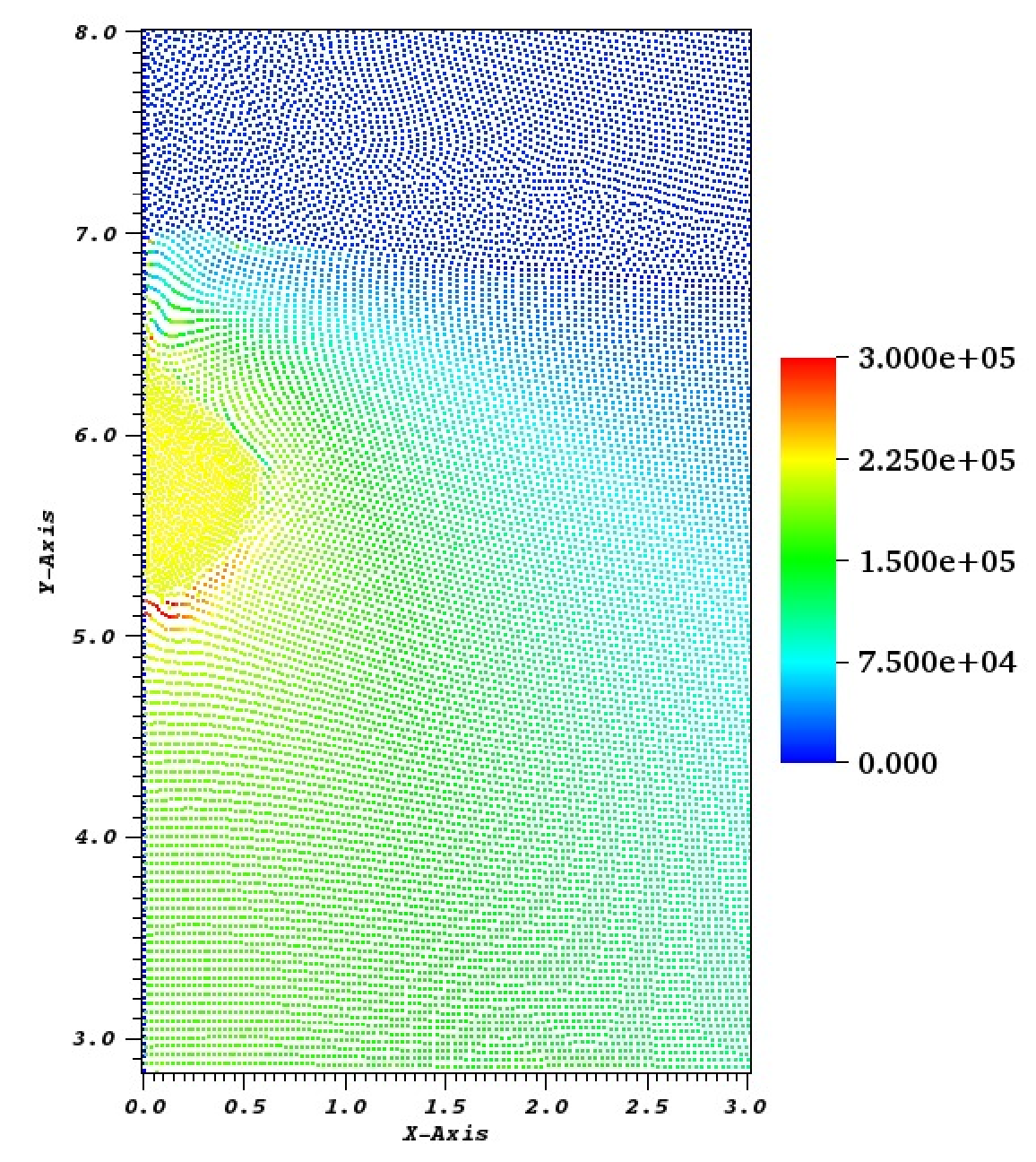}}
  \subfigure[]{\includegraphics[width=0.32\textwidth]{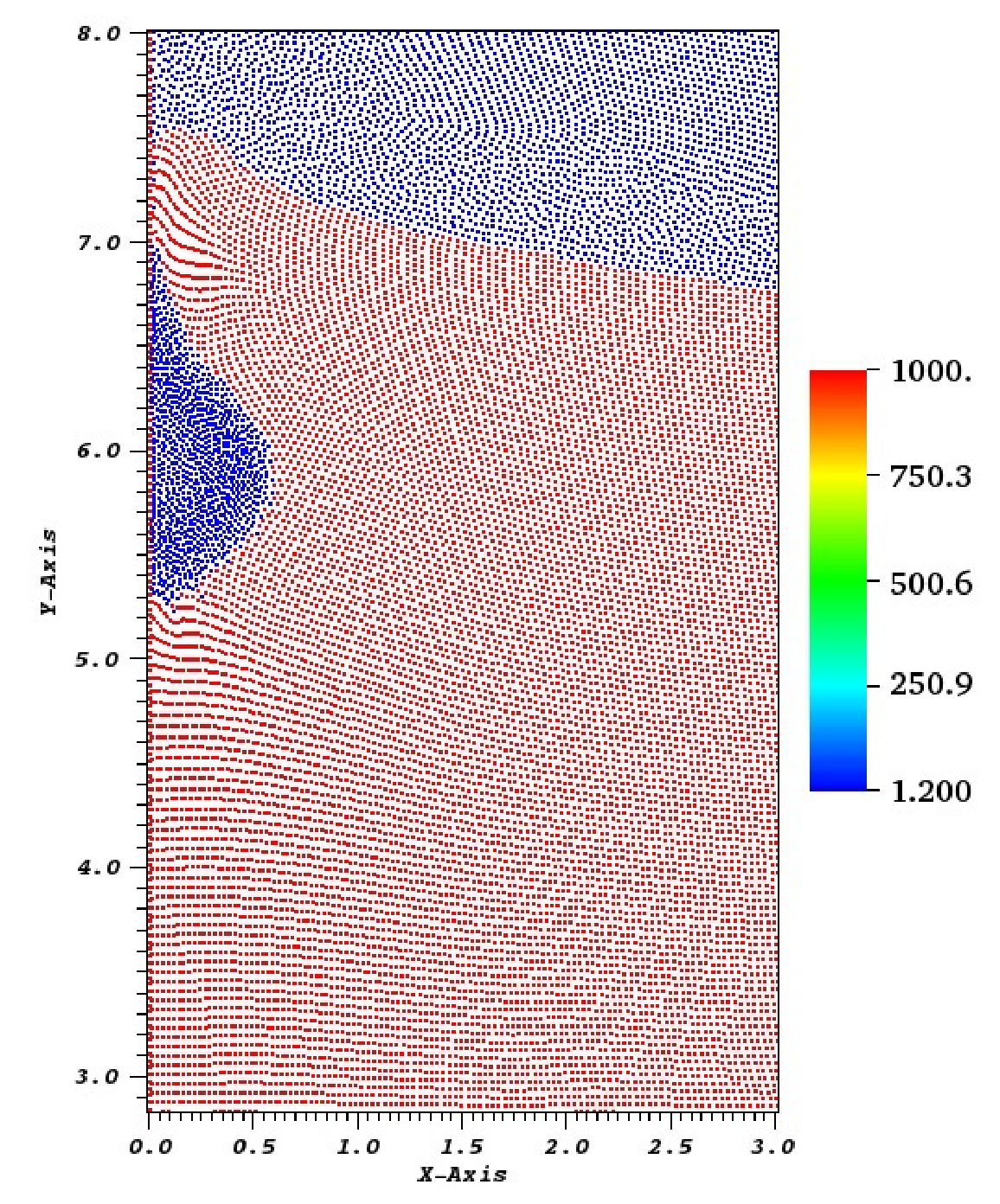}}
  \subfigure[]{\includegraphics[width=0.32\textwidth]{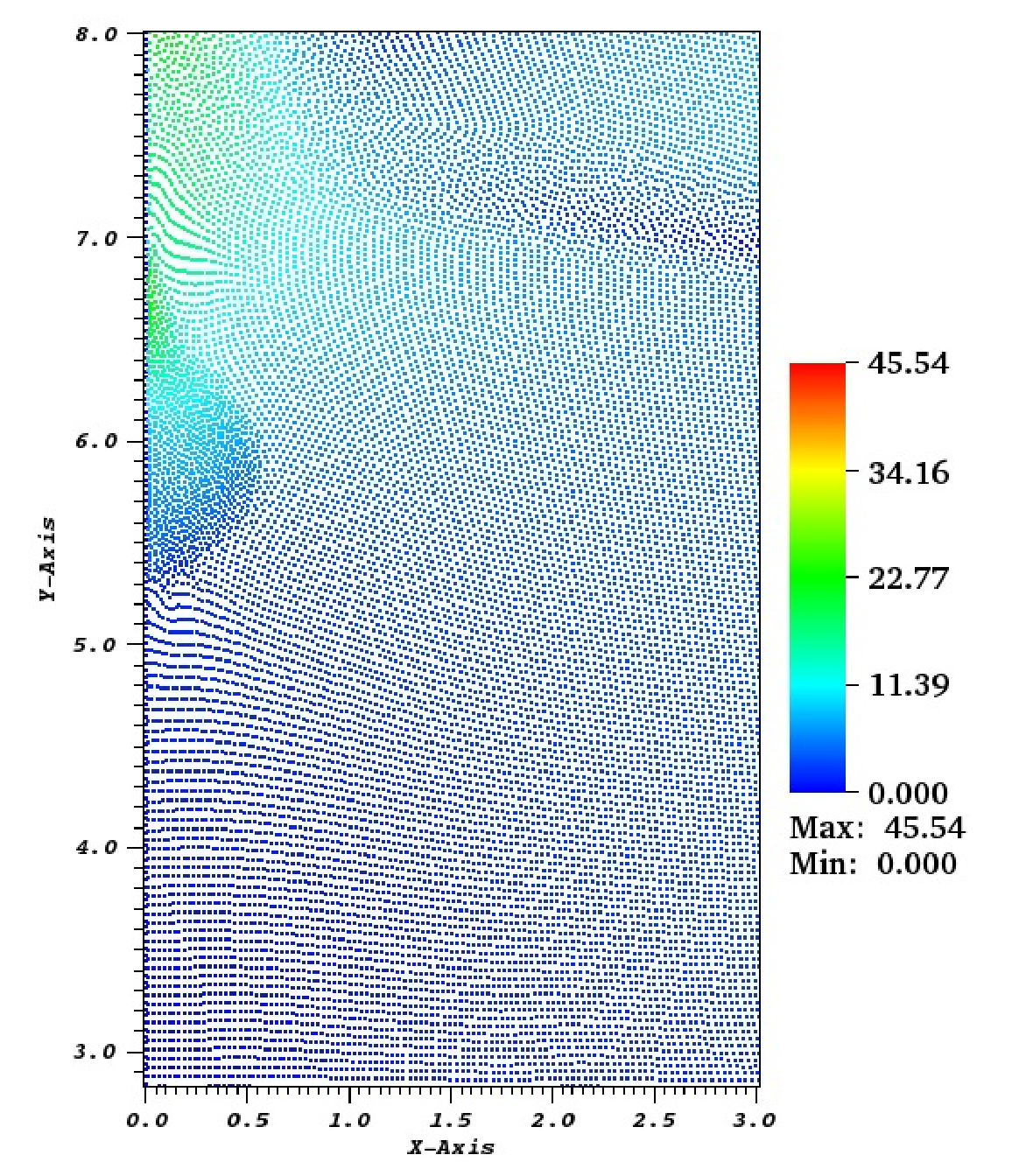}}
  \subfigure[]{\includegraphics[width=0.32\textwidth]{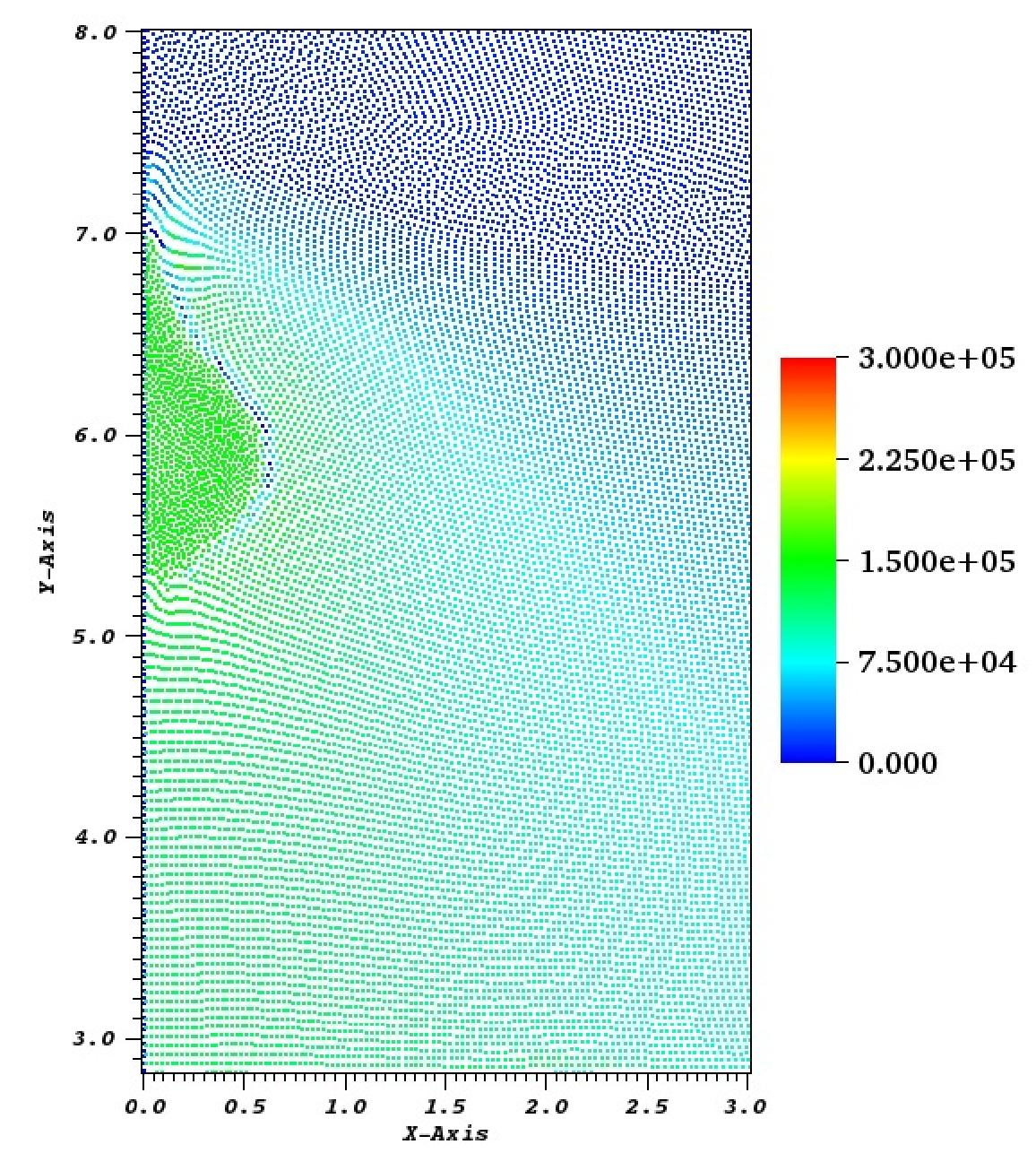}}
  \subfigure[]{\includegraphics[width=0.32\textwidth]{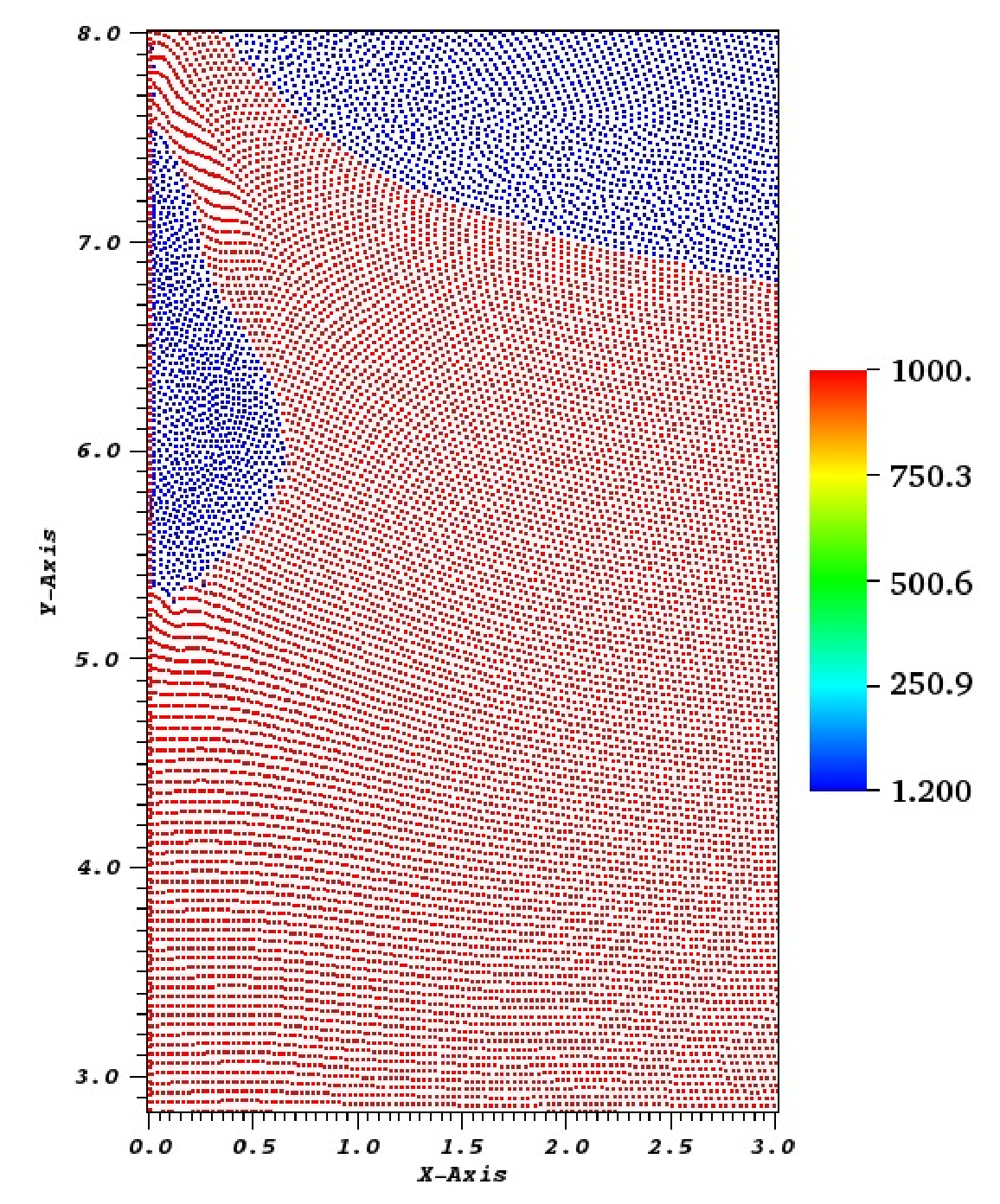}}
  \subfigure[]{\includegraphics[width=0.32\textwidth]{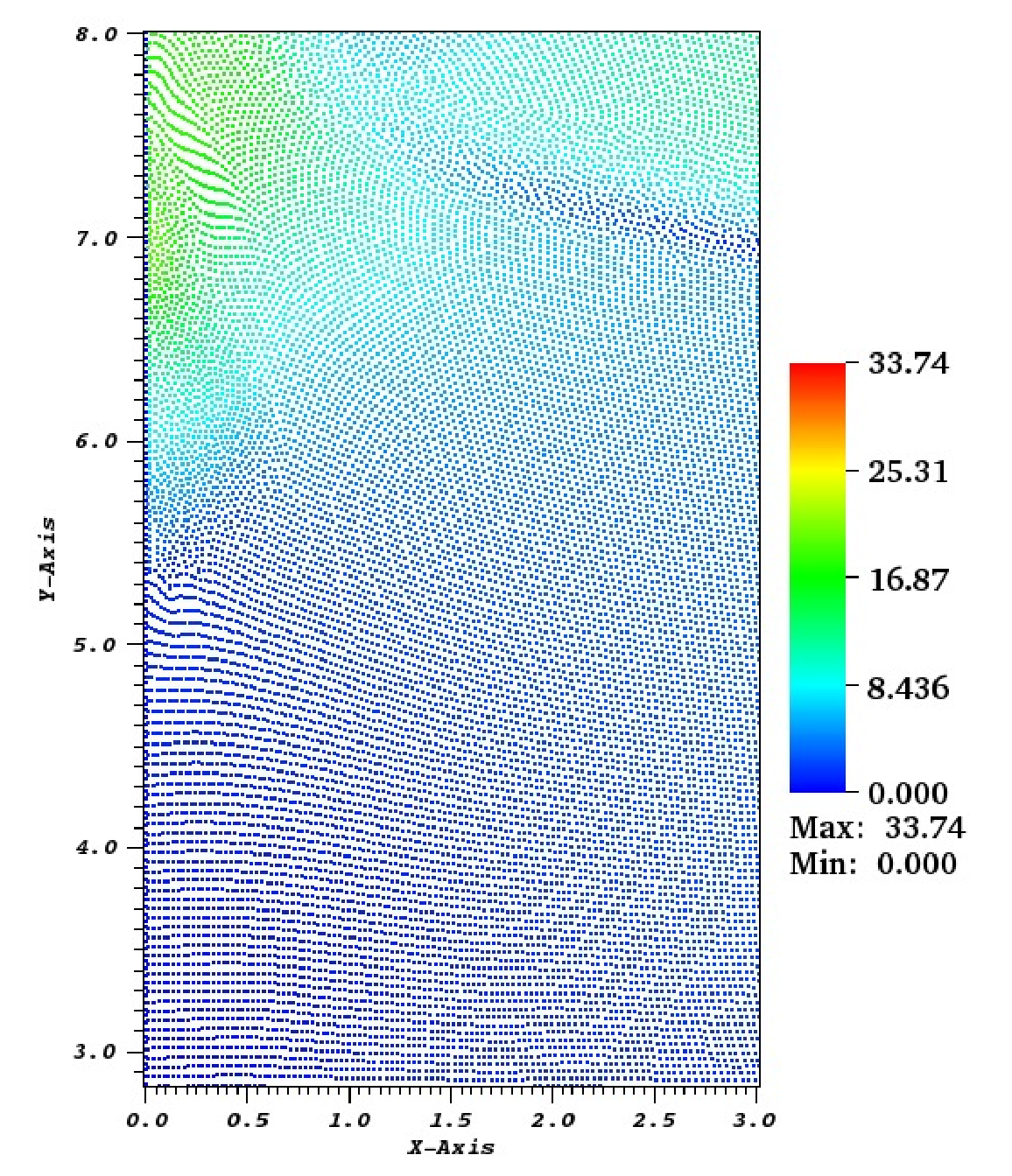}}
  \subfigure[]{\includegraphics[width=0.32\textwidth]{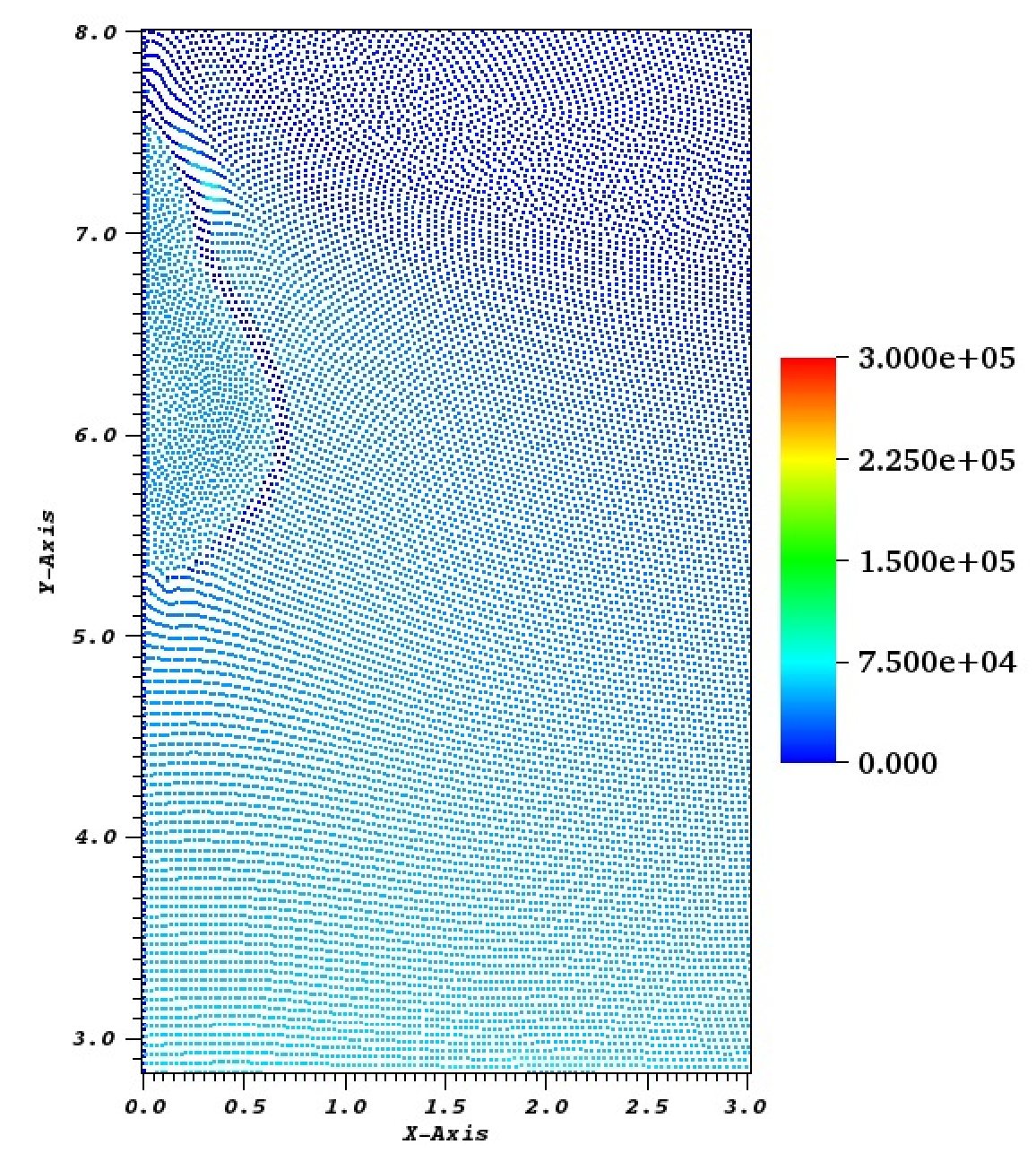}}
  \caption{\small\em Snapshots of the SPH simulations at various times. Particles are coloured by their density (left), velocity magnitude (middle) and pressure (right). Time increases from top to the bottom. Continued on the next page.}
  \label{sph_fsid_zoom2}
\end{figure}

The SPH particles were placed on a grid of squares with initial spacing of $\Delta p = 0.033$ $\mathsf{m}$ resulting in a total number of $\sim 154,000$ particles including the ghost boundaries. Figure~\ref{sph_fsid} shows the snapshots of the SPH particles at various times during the impact. In order to better understand the pressure distribution and the velocity field, closer views of the particles during the impact are shown in Figures \ref{sph_fsid_zoom} and \ref{sph_fsid_zoom2}.

Three main characteristic phenomena can be clearly seen in the figures. (\expandafter{\romannumeral 1}) As the wave crest is approaching the wall, the gas tends to escape between the wave crest and the wall. Figure~\ref{sph_fsid_zoom}(e)  shows the time when the escaping gas reaches its maximum velocity of $122.6~\mathsf{m}\,\mathsf{s}^{-1}$. (\expandafter{\romannumeral 2}) When the tip of the wave crest hits the wall, a maximum impact pressure occurs at the contact point. This is a very sharp and localised impact pressure (see Figure~\ref{sph_fsid_zoom}(i)). \textsc{Lafeber} \emph{et al.} (2012) \cite{Lafeber2012} introduced  and described the concept of Elementary Loading Processes (ELP) that occur during a single wave impact on a wall. They have coined this type of impact as an \textquotedblleft ELP1\textquotedblright~type impact. The ELP1-type impact is due to the discontinuity of velocity imposed by the wall to the liquid particles and characterised by instantaneously loaded area. This leads to a very sharp impact pressure peaks. (\expandafter{\romannumeral 3}) At this time the gas pocket is compressed and the pressure inside it oscillates. This type of pressure oscillations that are due to the gas compression and expansion are named as \textquotedblleft ELP3 \textquotedblright~in \cite{Lafeber2012}.

\begin{figure}
  \centering
  \input{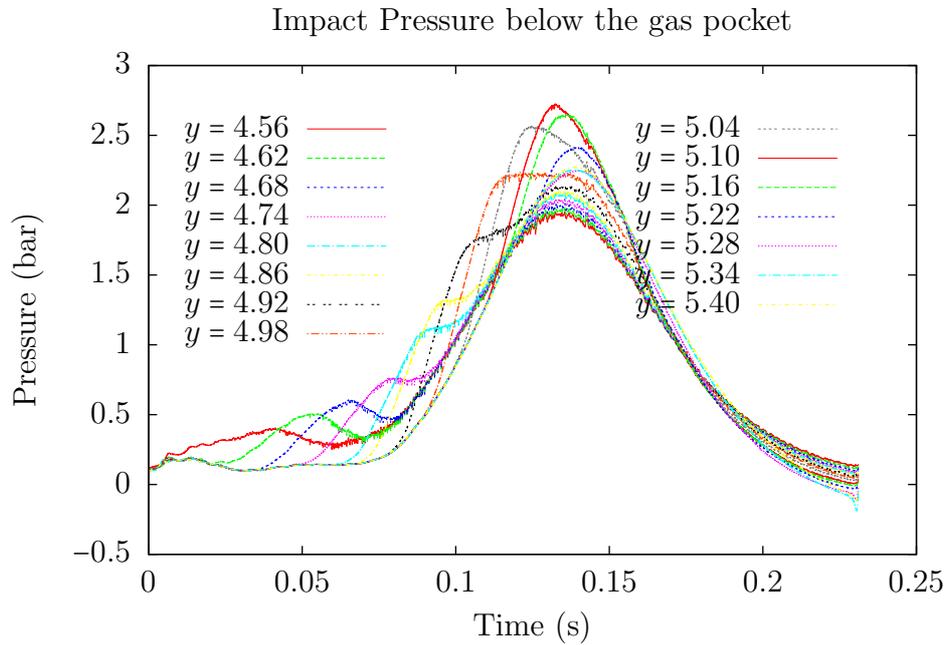}
  \caption{\small\em Time history of the pressure variation at pressure sensors initially located below the water level on the wall.}
  \label{impact1}
\end{figure}

\begin{figure}
  \centering
  \input{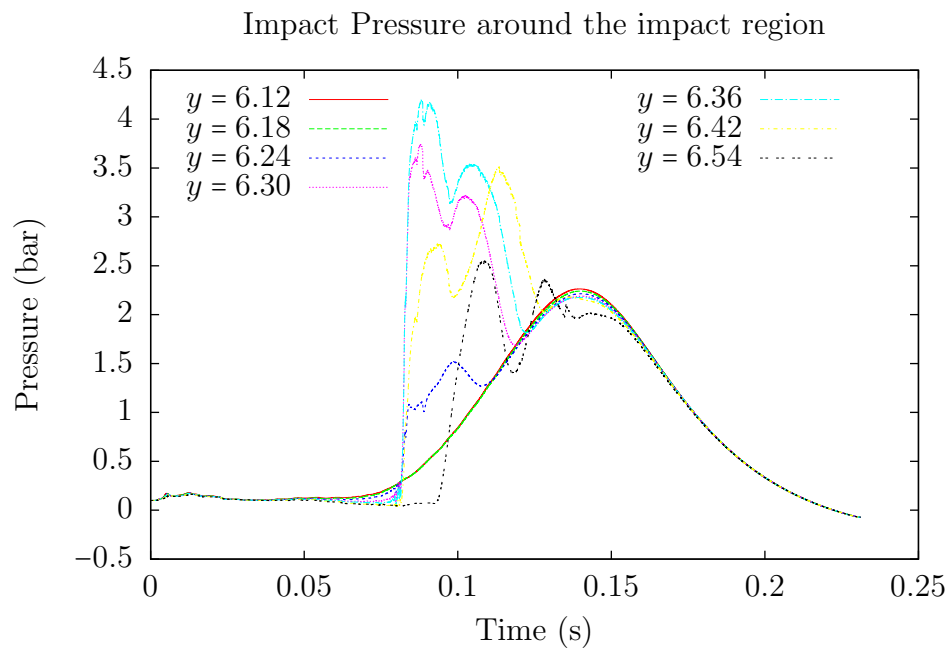}
  \caption{\small\em Time history of the pressure variation at pressure sensors located around the impact point.}
  \label{impact4}
\end{figure}

\begin{figure}
  \centering
  \input{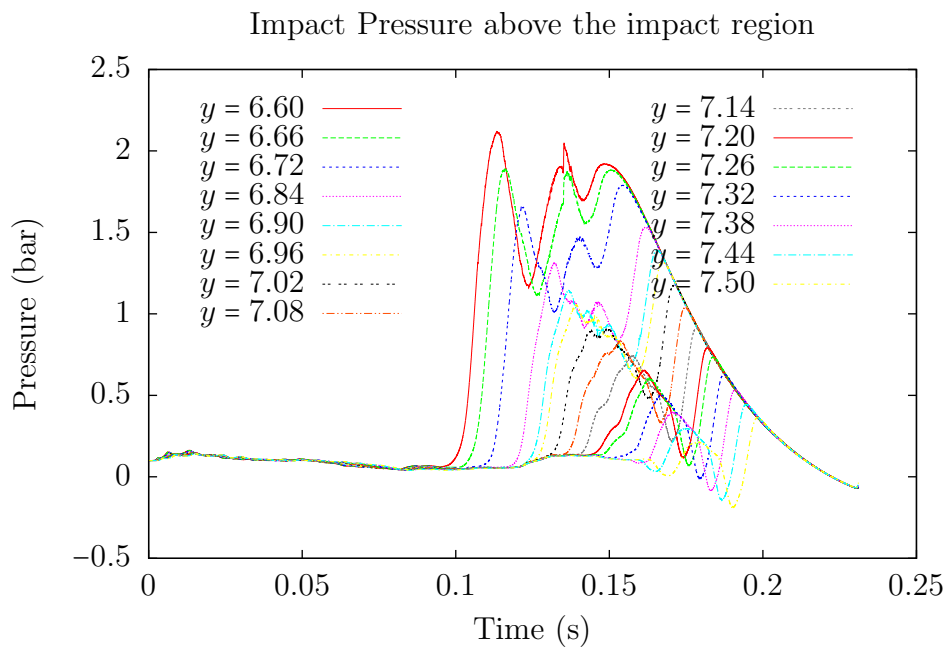}
  \caption{\small\em Time history of the pressure variation at pressure sensors located above the impact region.}
  \label{impact5}
\end{figure}

When the wave gets closer to the wall, the water level at the wall is gradually increasing at the points initially below it. This increase in the free-surface level results in a slight increase of pressure in this region. Figure~\ref{impact1} shows the time history of the pressure variations at these points. The pressure inside the gas pocket is however smooth and uniform in time as observed by \cite{Lafeber2012, Guilcher2012, Costes2013}. The time variation of the pressure for sensors located inside the gas pockets is shown in Figure~\ref{impact1}. For the sensors located at the impact region, the impact pressure is very sharp and hence requires a very fine resolution in both space and time. Figure~\ref{impact4} shows the impact pressure profile at the sensors around the impact location. It can be seen from the sensors located at $y = 6.24$ $\mathsf{m}$ and at $y = 6.30$ $\mathsf{m}$ that the peak of the impact pressure is four times larger. Since the difference between these two sensors is less than two particles spacing, it shows that a much finer resolution is required to be able to capture the impact peak precisely. Figure~\ref{impact4} also shows the decay of the pressure at the sensors away from the impact point. As the wave travels along the wall after the impact, the pressure increases on the sensors that get into contact with the run-up and decreases afterwards. This can be clearly seen in Figure~\ref{impact5}. However, as the wave travels along the wall it looses its momentum and hence the peak of the travelling pressure is reduced.

\begin{figure}
  \centering
  \input{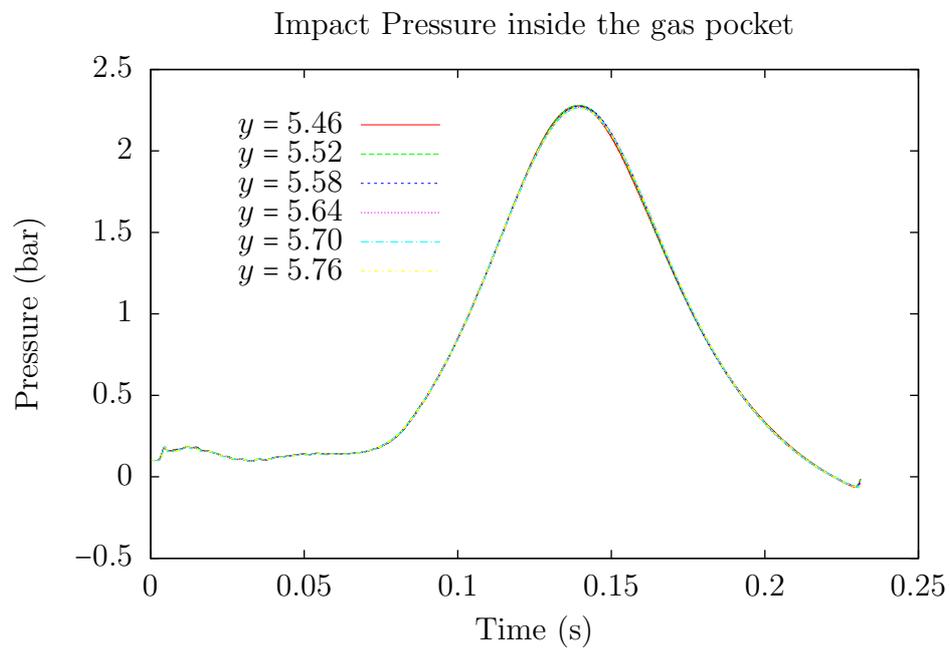}
  \caption{\small\em Time history of the pressure variation at pressure sensors located inside the gas pocket.}
  \label{impact2}
\end{figure}

\section{Conclusion}\label{sec:conclude}

Simulating two-phase compressible flows is known to be challenging with the SPH method. In addition, modelling a quasi--incompressible flow with a compressible solver is a cumbersome problem regardless of the nature of the CFD approach. In this paper, a robust, reliable and accurate two-phase SPH solver is introduced to model complex fluid dynamics problems.

Two different problems are solved to demonstrate the capability of the two-phase SPH scheme. The first test case is an interesting benchmark problem as it incorporates the phenomenology of liquid impact during complex events such as sloshing in marine tanks. The SPH results were compared with the results of an incompressible Level Set solver and showed an excellent agreement in terms of the pressure peak at the same resolutions with the Level Set simulations. The difference in the temporal location of the impact pressure was associated to the interface thickness in the Level Set simulations whereas the proposed SPH scheme was capable of modelling very sharp interfaces between the phases.

The second test case involved a wave impact on a rigid wall with an entrained gas pocket. Here, the SPH particles were initialised by the output of a fully non-linear potential flow solver (FSID) \cite{Scolan2010}. Although the global features of the flow were modelled very well, a much finer resolution is required to capture the impact pressure accurately. The parallelization of the code is in progress and the simulations with finer resolutions will be performed in the nearest future.

\subsection*{Acknowledgments}
\addcontentsline{toc}{section}{Acknowledgments}

This work has been funded by Science Foundation Ireland (SFI) under the research project ``High--end computational modelling for wave energy systems''. The authors would like to thank the Irish Centre for High--End Computing (ICHEC) for the provision of computational facilities and support. 

This work was granted access to the HPC resources of the Swiss National Supercomputing Centre (CSCS)/Mount Rosa -- Cray XE6 cluster made available within the Distributed European Computing Initiative by the PRACE--2IP, receiving funding from the European Community's Seventh Framework Programme (FP7/2007-2013) under grant agreement n$^\circ$ RI--283493.

D.~\textsc{Dutykh} and F.~\textsc{Dias} would also like to acknowledge the support from ERC under the research project ERC-2011-AdG 290562-MULTIWAVE.

\addcontentsline{toc}{section}{References}
\bibliographystyle{abbrv}
\bibliography{biblio}

\end{document}